\documentstyle[prb,epsf,floats,aps,amsfonts,twocolumn,eqsecnum]{revtex}

\input epsf
\newwrite\ffile\global\newcount\figno \global\figno=1

\def\writedef#1{}

\def\figin{\epsfcheck\figin}\def\figins{\epsfcheck\figins}
\def\epsfcheck{\ifx\epsfbox\UnDeFiNeD
\message{(NO epsf.tex, FIGURES WILL BE IGNORED)}
\gdef\figin##1{\vskip2in}\gdef\figins##1{\hskip.5in}
\else\message{(FIGURES WILL BE INCLUDED)}%
\gdef\figin##1{##1}\gdef\figins##1{##1}\fi}

\def\figinsert{}
\def\ifig#1#2#3{\xdef#1{fig.~\the\figno}
\writedef{#1\leftbracket fig.\noexpand~\the\figno}%
\figinsert\figin{\centerline{#3}}
\medskip\centerline{\vbox{\baselineskip12pt
\advance\hsize by -1truein\center\small{Fig.~\the\figno.} #2}}
\bigskip\endinsert\global\advance\figno by1}
\def\endinsert{}

\newcommand{\beq}{\begin{equation}}
\newcommand{\eeq}{\end{equation}}
\newcommand{\bea}{\begin{eqnarray}}
\newcommand{\eea}{\end{eqnarray}}
\newcommand{\no}{\nonumber}
\newcommand{\br}{\no\\&&}

\newcommand{\ad}{a^{\dagger}}
\newcommand{\bd}{b^{\dagger}}
\newcommand{\fd}{f^{\dagger}}

\newcommand{\cd}{c^{\dagger}}
\newcommand{\sd}{s^{\dagger}}

\newcommand{\disave}[1]{\bigl[ #1 \bigr]}

\newcommand{\aph}{a^{\vphantom{\dagger}}}
\newcommand{\bph}{b^{\vphantom{\dagger}}}
\newcommand{\fph}{f^{\vphantom{\dagger}}}

\newcommand{\cph}{c^{\vphantom{\dagger}}}
\newcommand{\sph}{s^{\vphantom{\dagger}}}

\newcommand{\ra}{\rangle}
\newcommand{\la}{\langle}

\newcommand{\dx}{\partial_x}

\newcommand{\os}{{osp($2n\,|\,2n$)}}
\newcommand{\osp}{{osp($2n+1\,|\,2n$)}}
\newcommand{\Osp}{{osp($2n+2\,|\,2n$)}}

\newcommand{\ab}{{\bar a}}
\newcommand{\bb}{{\bar b}}
\newcommand{\fb}{{\bar f}}
\newcommand{\eb}{{\bar e}}
\newcommand{\hb}{{\bar h}}

\newcommand{\bi}{\bbox{i}}
\newcommand{\bj}{\bbox{j}}
\newcommand{\bx}{\hat{\bbox{x}}}
\newcommand{\by}{\hat{\bbox{y}}}

\newcommand{\nb}{n_b}

\newcommand{\nf}{n_f}
\newcommand{\nbb}{n_\bb}
\newcommand{\nfb}{n_\fb}

\newcommand{\al}{\alpha}
\newcommand{\be}{\beta}
\newcommand{\ga}{\gamma}
\newcommand{\de}{\delta}

\newcommand{\eps}{\epsilon}

\newcommand{\om}{\omega}
\newcommand{\lam}{\lambda}

\newcommand{\La}{\Lambda}
\newcommand{\Si}{\Sigma}

\newcommand{\sx}{\sigma^x}
\newcommand{\sy}{\sigma^y}
\newcommand{\sz}{\sigma^z}

\newcommand{\abd}{\ab^{\dagger}}
\newcommand{\bbd}{\bb^{\dagger}}
\newcommand{\fbd}{\fb^{\dagger}}

\newcommand{\abph}{\ab^{\vphantom{\dagger}}}
\newcommand{\bbph}{\bb^{\vphantom{\dagger}}}
\newcommand{\fbph}{\fb^{\vphantom{\dagger}}}

\newcommand{\Jab}{J^{\vphantom{\dagger}}_{\al \be}}
\newcommand{\jab}{J_{\al \be}}
\newcommand{\dab}{\de_{\al \be}}

\newcommand{\Nfi}{{\hat N}_{Fi}}
\newcommand{\Nbi}{{\hat N}_{Bi}}
\newcommand{\NSi}{{\hat N}_{Si}}

\newcommand{\Xfi}{{\hat X}_{Fi}}
\newcommand{\Xbi}{{\hat X}_{Bi}}
\newcommand{\XSi}{{\hat X}_{Si}}

\newcommand{\Tr}{\,\text{Tr}\,}
\newcommand{\STr}{\,\text{STr}\,}
\newcommand{\str}{\,\text{str}\,}

\newcommand{\Tt}{{\rm T}_{\tau}}

\begin{document}

\twocolumn[\hsize\textwidth\columnwidth\hsize\csname
@twocolumnfalse\endcsname

\title{Random-bond Ising model in two dimensions: The Nishimori line
and supersymmetry}

\author{Ilya A. Gruzberg$^1$, N. Read$^2$, and Andreas W. W. Ludwig$^3$}
\address{$^1$Institute for Theoretical Physics, University of California,
Santa Barbara, CA 93106-4030 \\
$^2$Departments of Physics and
Applied Physics, Yale University, P.O. Box 208120, New Haven, CT
06520-8120\\
$^3$Physics Department, University of California,
Santa Barbara, CA 93106-9530}

\date{July 14, 2000}

\maketitle

\begin{abstract}
We consider a classical random-bond Ising model (RBIM) with binary
distribution of $\pm K$ bonds on the square lattice at finite
temperature. In the phase diagram of this model there is the
so-called Nishimori line which intersects the phase boundary at a
multicritical point. It is known that the correlation functions
obey many exact identities on this line. We use a supersymmetry
method to treat the disorder. In this approach the transfer
matrices of the model on the Nishimori line have an enhanced
supersymmetry {\osp}, in contrast to the rest of the phase
diagram, where the symmetry is {\os} (where $n$ is an arbitrary
positive integer). An anisotropic limit of the model leads to a
one-dimensional quantum Hamiltonian describing a chain of
interacting superspins, which are irreducible representations of
the {\osp} superalgebra. By generalizing this superspin chain, we
embed it into a wider class of models. These include other models
that have been studied previously in one and two dimensions. We
suggest that the multicritical behavior in two dimensions of a
class of these generalized models (possibly not including the
multicritical point in the RBIM itself) may be governed by a
single fixed point, at which the supersymmetry is enhanced still
further to {\Osp}. This suggestion is supported by a calculation
of the renormalization-group flows for the corresponding nonlinear
sigma models at weak coupling.

\end{abstract}

 ]

\section{Introduction}

For many decades Ising models served as the simplest nontrivial
models for the description of magnetically ordered phases and
phase transitions between them. This is true both for pure models
and for Ising models with randomness. In particular, in the
context of the spin glass problem  \cite{by} the relevant Ising
models have random bonds of both signs (ferro and
antiferromagnetic). This leads to frustration and the possibility
of spin glass order.

In this paper we consider a classical random-bond Ising model
(RBIM) of Ising spins $S_{\bi} = \pm 1$ on the two-dimensional
(2D) square lattice with the Hamiltonian ($\be=1/T$ is the inverse
temperature) \beq \be{\cal H} = - \sum_{\la\bi\bj\ra} K_{\bi\bj}
S_{\bi} S_{\bj}, \label{ham} \eeq where the bold indices $\bi =
(i_x, i_y)$ and $\bj = (j_x, j_y)$ denote 2D vectors of integer
coordinates of the sites of the lattice, the summation is over
distinct nearest-neighbor bonds (i.e.\ pairs), and the coupling
constants $K_{\bi\bj}$ are independent random variables drawn from
the distribution \beq P[K_{\bi\bj}] = (1-p)\de(K_{\bi\bj} - K) +
p\de(K_{\bi\bj} + K). \label{PK} \eeq In words, the couplings
$K_{\bi\bj}$ are ferromagnetic ($K > 0$) with probability $1-p$
and antiferromagnetic with probability $p$. Notice that $K$ varies
inversely with $T$. In what follows we will occasionally also
consider Ising models with other distributions of the bond
strengths. For simplicity, in most cases, where it cannot lead to
confusion, we will simply call the model with the binary
distribution (\ref{PK}) ``the RBIM''. Later, we will also consider
the anisotropic generalization of the model, in which $K$ takes
different values on bonds in the $x$ and $y$ directions.

\begin{figure}
\epsfxsize=3.0in \vspace{-1mm}
\centerline{\epsffile{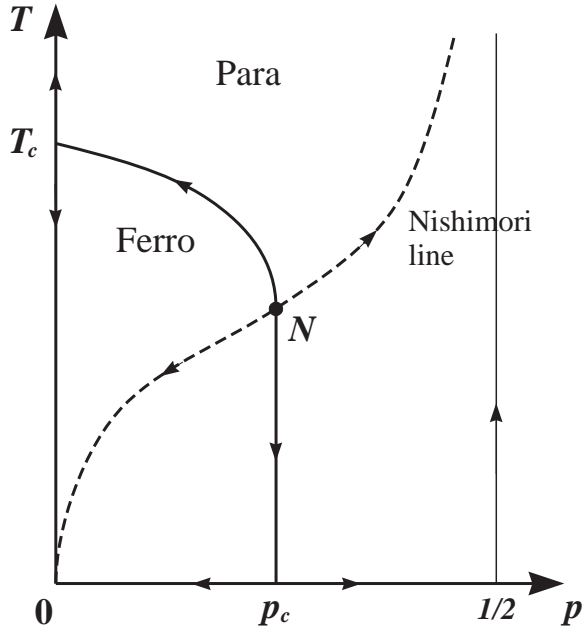}} \vspace{0.5cm}
\caption[Phase diagram of the random-bond Ising model]{Phase
diagram of the random-bond Ising model of Eq.\
(\protect\ref{ham}), in terms of $T \propto 1/K$ and $p$.}
\label{fig1}
\end{figure}

Let us summarize some of what is known about this model. The phase
diagram of this model is still somewhat controversial, but is
widely believed to be as in Fig.\ \ref{fig1}
\cite{phasediag,dot,n,vert,ghld,ldgh,arqs}. First we note that for
$p=1$, we have a pure antiferromagnetic Ising model, which can be
mapped onto the ferromagnetic case by sending $S_{\bi}\to
-S_{\bi}$ for $\bi$ on one sublattice. More generally, this
transformation is equivalent to sending $p\to 1-p$. Hence we need
show only the region $0\leq p \leq 1/2$. The solid line is a phase
boundary which separates the ferromagnetically-ordered from the
paramagnetic phase. Fig.\ \ref{fig1} can also be viewed as a
schematic renormalization group (RG) flow diagram, in which the
intersection points labeled $T_c$ (corresponding to
$K_c=0.44\ldots$, the pure Ising transition), $p_c\simeq 0.12$
(the $T=0$ transition), and another point $N$ are viewed as RG
fixed points that govern the critical behavior for the portions of
the phase boundary shown as flowing into these points ($N$ is an
unstable, hence multicritical, point). In 2D, it is generally
believed that no spin glass phase exists at finite temperature. At
zero temperature, long-range spin-glass (Edwards-Anderson) order
exists trivially when the distribution of bonds is continuous,
since in a finite system there is, with probability one, a unique
ground state, up to a reversal of all the spins. (Taking the
thermodynamic limit in a fixed sample is a very subtle problem;
for a very recent discussion, see Ref.\ \onlinecite{ns} and
references therein.) However, for the discrete distribution with
bonds taking values $K$, $-K$, assumed here, the existence of such
order in the region $p>p_c$ is not clear, because there will be
many degenerate ground states. There is evidence for power-law
spin-glass correlations at $T=0$ in this region \cite{power,bp}.
In three or more dimensions, there is a spin-glass ordered phase
at temperatures below some temperature $T_{\rm SG}(p)$, and which
extends up to $p=1/2$. The three phase boundaries meet in a
multicritical point at some $T<T_c$, $p<1/2$. The point $N$ in 2D
is in some sense a remnant of this multicriticality in higher
dimensions.

An interesting feature of this phase diagram is the so-called
Nishimori line (NL), shown dashed. Such a line can be defined for
a broad class of distributions $P[K_{\bi\bj}]$, and in our special
case (\ref{PK}) is given by the condition \beq 1-2p = \tanh K.
\label{nishimori} \eeq Nishimori \cite{n} found that the model on
the NL has a local (``gauge'') ${\Bbb Z}_2$ symmetry. Using the
symmetry he showed that the internal energy of the RBIM is
analytic along the NL, and also established a special case of the
following identities for correlation functions (proven generally
in Ref. \onlinecite{ldgh}) which hold on the NL \beq \disave{\la
S_{\bi_1} \ldots S_{\bi_k}\ra^{2q-1}} = \disave{\la S_{\bi_1}
\ldots S_{\bi_k}\ra^{2q}}. \label{correq} \eeq Here
$\{S_{\bi_1},\ldots, S_{\bi_k}\}$ is any set of the Ising spins,
the angular brackets denote the thermal average, and the square
brackets the average over the distribution of bonds $P[K]$.
Georges {\it et al\/} \cite{ghld} showed that Nishimori's result
concerning the internal energy may be rederived using a
supersymmetric formulation.  Nishimori later argued, and Kitatani
showed \cite{vert}, that the ferro-para phase boundary is vertical
below the NL.  This is supported by the later numerical works
among Refs.\ \onlinecite{phasediag,arqs}.

Le Doussal, Georges and Harris \cite{ldgh} established that in any
dimension the NL goes through the  multicritical point $N$ on the
ferro-para phase boundary and is one of the exact RG trajectories
near this point, the other trajectory being the vertical tangent
to the phase boundary at $N$. The RG eigenvalues along these
trajectories and the corresponding critical exponents were
estimated by Singh and Adler \cite{sa} from high-temperature
expansions of high order. Two of their exponents in 2D are very
close to those of classical 2D percolation, as are the results in
Ref.\ \onlinecite{arqs}. We emphasize that, while Nishimori's
results do not apply to every possible distribution of disorder in
the RBIM, the implications for universal critical properties must
hold throughout a universality class, and it is believed that this
class includes the generic RBIM multicritical point \cite{ldgh}.

More recently, Cho and Fisher \cite {cf} proposed a network model
similar to the Chalker-Coddington model used to describe
transitions between integer quantum Hall plateaus \cite{cc}. The
Cho-Fisher model was supposed to be in the same universality class as the
2D RBIM. They simulated their network numerically and found a phase
diagram with a multicritical point, with numerical critical exponents
close to the ones found by Singh and Adler \cite{sa}. In her thesis
\cite{chothes}, Cho also simulated another network model which is
precisely equivalent to the RBIM, and again found exponents close to some
of those in Ref.\ \onlinecite{sa}.

Here we will briefly sketch a version of the argument of Ref.\
\onlinecite{ldgh}, since it provided motivation for our work. Let
us consider the random-bond Ising model with Gaussian disorder,
\beq P[K_{\bi\bj}]=\frac{1}{\sqrt{2\pi\Delta}}
\exp[-(K_{\bi\bj}-K_0)^2/(2\Delta^2)], \eeq so the mean of
$K_{\bi\bj}$ is $K_0$, and the standard deviation is $\Delta$.
Taking the partition function \beq
Z=\sum_{\{S_{\bi}\}}\exp\sum_{\la\bi\bj\ra}K_{\bi\bj}S_{\bi}
S_{\bj}, \eeq we replicate and average to obtain \beq
[Z^n]=\sum_{\{S_{\bi}^a\}}\exp\left[K_0\sum_{\la\bi\bj\ra,a}S_{\bi}^a
S_{\bj}^a +\frac{1}{2}\Delta^2\sum_{\la\bi\bj\ra,ab}S_{\bi}^a
S_{\bj}^a S_{\bi}^b S_{\bj}^b\right], \eeq where $a$, $b=1$,
\ldots, $n$. For finite $n$ this has the form of the Ashkin-Teller
model, consisting of $n$ coupled Ising models. Now compare this
with the replicated spin-glass model, which is obtained by setting
$K_0=0$: \beq [Z^m]=\sum_{\{S_{\bi}^a\}}\exp\left[
\frac{1}{2}\Delta^2\sum_{\la\bi\bj\ra,ab}S_{\bi}^a S_{\bj}^a
S_{\bi}^b S_{\bj}^b\right], \eeq with $a$, $b=0$, \ldots, $m-1$.
This model has a gauge symmetry: it is invariant under
site-dependent transformations $S_{\bi}^a\to-S_{\bi}^a$ for all
$a$ and any set of $\bi$'s. This local ${\Bbb Z}_2$ gauge symmetry
can be fixed by setting all $S_{\bi}^a=1$ for all $\bi$, for one
value of $a$, say $a=0$. Then if $m=n+1$, we obtain the
random-bond partition function with $K_0=\Delta^2$, up to
constants. On this line, which is the NL for the Gaussian case,
there are still consequences of the underlying gauge symmetry. The
$m=n+1$ replicas are still on an equal footing, and this leads to
the identities (\ref{correq}) on the NL, as follows. In the
gauge-unfixed model, only correlation functions containing an even
number of $S_{\bi}^a$'s at each site are nonzero. The correlation
functions are invariant under permutations of the replicas, and so
independent of whether or not $a=0$ is among the components. Since
different replica components represent distinct thermal averages
in the $n \to 0$ limit, we obtain the identities (\ref{correq}),
and others, on gauge fixing. The ${\Bbb Z}_2$ local gauge symmetry
in the replica formalism should not be confused with that of
Nishimori, who did not use replicas; it is the enlarged
permutational symmetry of the replicas that corresponds to
Nishimori's arguments. Off the NL, the identities are lost, but
the model can still be written as a gauge-fixed version of a
system with $n+1$ replicas and a local ${\Bbb Z}_2$ gauge
symmetry.

The preceding argument shows that, in the replica formalism, the
NL is special because it possesses a larger permutation symmetry
S$_{n+1}$ in place of the usual S$_n$. As we saw, even off the NL,
an additional ``zeroth'' replica spin can be introduced into the
model, along with a gauge symmetry that can be used to remove the
unwanted degrees of freedom, and further {\em on} the NL line the
zeroth spin is symmetric with the others. Now from work extending
back to Onsager, the 2D Ising model can be written in terms of
free fermions, which become Majorana (real Dirac) fermions in the
continuum limit, and furthermore Ashkin-Teller models can be
represented by interacting Majorana fermions, with O($n$) symmetry
\cite{shank85}. Hence we are led to conjecture that, in the
replicated fermion representation, it is possible to introduce an
additional ``zeroth'' fermion, together with a local ${\Bbb Z}_2$
gauge symmetry to remove the unwanted degrees of freedom, and that
on the NL, we should find a larger O($n+1$) symmetry. In this
paper we demonstrate that this indeed occurs, though we take a
different route to do so. We consider the binary distribution
above, and we use supersymmetry rather than replicas, so no limit
$n \to 0$ need be taken. However, the corresponding result for
replicas is contained in our results. The network models such as
the Cho-Fisher model also depend on the fermion representation of
the Ising model, and so we can also consider these models in our
framework. We find that the models can be viewed as supersymmetric
vertex models, or by using the anisotropic limit as Hamiltonian
chains, which act in irreducible representations of the relevant
symmetry (super-)group (which is enlarged on the NL), and thus are
quantum spin chains, and possibly can also be viewed as the
strong-coupling region of a nonlinear sigma model. This greatly
enhances the similarity of the problem to the integer quantum Hall
effect transition, and to other random fermion problems. However,
we find that the NL does not fall into a recent list of nonlinear
sigma models that correspond to random matrix ensembles in such
problems \cite{az}. Our results also apply to certain
one-dimensional fermion problems.

In random fermion problems, including those arising in disordered
superconductors, it is usual to attempt classifications, based on
symmetries, for {\em generic} probability distributions, as in
Ref.\ \onlinecite{az} for random matrices. The ensembles found in
Ref.\ \onlinecite{az} for disordered superconductors differ from
the standard ensembles because of the lack of a conserved particle
number, and because zero energy is a special point in the
spectrum. The second-quantized noninteracting quasiparticle
description can in each case be replaced by a ``first-quantized''
formalism involving a single matrix, which must satisfy certain
discrete symmetry and symmetry-like conditions, which distinguish
the ensembles. One such class, termed class D in Ref.\
\onlinecite{az}, is for problems with broken time reversal and no
spin-rotation symmetry. This corresponds to the symmetries of the
fermion representation of the RBIM. In this class, a nonlinear
sigma model analysis in two dimensions indicates that there is a
metallic phase in which the fermion eigenstates are extended
\cite{bundschuh,sfnew,readgr,bsz}. Senthil and Fisher \cite{sfnew}
discussed a scenario in which such a phase occurs in the RBIM, as
an intermediate phase between the para- and ferromagnetic phases,
at low $T$ in the region labeled ``Para'' in Fig.\ \ref{fig1}, and
with its bordering phase boundaries (one of which is the low $T$
phase boundary shown) meeting at the multicritical point $N$ on
the NL. (This is the region sometimes claimed to be some kind of
spin-glass--like phase in the RBIM literature.) They suggested
that the phase would be characterized by the absence of long range
order in the mean of either the ferromagnetic Ising or the dual
disorder variable correlations \cite{sfnew}. Presumably such decay
would also hold for the mean square (spin-glass) correlations. It
is not clear if this is consistent with the $T\to 0$ analysis of
Ref.\ \onlinecite{bp}. An alternative scenario is that a finer
analysis of problems with broken time-reversal symmetry is needed,
and that the nonlinear sigma model appropriate for class D does
not apply to the RBIM. Indeed, a recent paper \cite{bsz}
emphasizes that the target manifold of the class D model is not
connected, and that consequently there can be domains of the two
components or ``phases''. It follows that additional parameters
are required in order to fully parametrize the systems,
independent of those familiar for sigma models with connected
targets. This allows for a much richer phase diagram and
transitions in this symmetry class. We show in this paper that
this is connected with the structure we uncover in the random-bond
Ising and network models. In another paper \cite{rl}, it is argued
that the metallic phase cannot occur in a RBIM with real Ising
couplings.

Problems of random noninteracting fermions are among the better
understood of disordered systems, and while many results are
numerical, in some cases there are even exact results for critical
properties in 2D. By making contact between the RBIM and other
random fermion problems, and casting them in a common language, we
hope to gain understanding of this disordered classical spin
problem. At the same time, the RBIM provides an example that may
shed light on previously-unknown classes or ensembles of random
fermions. The analysis presented in this paper does not resolve
all aspects of the broken-time-reversal symmetry class, but it
does show that the NL is a special subclass.

We now give an outline of the paper. Sections \ref{tmatr},
\ref{enhsusy}, \ref{hamlimit} and \ref{general} contain the main
technical work. They show how a fermion representation \cite{sml}
can be used for the RBIM, and how bosons are also introduced to
cancel the inverse partition functions, via supersymmetry (SUSY)
\cite{c,fss,read,z,bf,km,sf1,glr}. The bosons live in a space with
an indefinite metric, a common feature of SUSY methods (see Refs.\
\onlinecite{read}, \onlinecite{glr}, and compare Refs.\
\onlinecite{bf,km,sf1}). On the NL, a larger SUSY algebra is
found. As an application of this enhanced SUSY, we use it in
Appendix \ref{eqcorr} to rederive the equalities (\ref{correq}).
An anisotropic limit of the model relates it to a Hamiltonian for
``superspins'' on ``split'' sites, two for each original site. The
Hamiltonian on the NL can be generalized in a natural way, and we
introduce a phase diagram for the generalized models; on one line
in this diagram the Hamiltonian possesses a still larger SUSY.
Section \ref{lowtemp} discusses a 1D model
\cite{bf,mckenzie,dyson,1d,bcgl}, and shows that it possesses a
single ``Nishimori'' point of higher SUSY. In Sec.\ \ref{CFm} we
consider the model of Cho and Fisher \cite{cf}, and show that it
does not correspond precisely to the RBIM, and nor does it have
the larger SUSY of the NL. In Sec.\ \ref{nlsm}, we argue that
much, or possibly all, of the critical surface in the phase
diagram in the higher SUSY generalized models is in the
universality class of the point with even higher SUSY mentioned
above. This is supported by consideration of nonlinear sigma
models that correspond to the spin chains, and a weak-coupling
calculation for these shows a renormalization-group flow towards
the higher SUSY theory. We comment on general network models and
nonlinear sigma models with the symmetries of class D. Appendix
\ref{reposp} gives details of a representation of SUSY that we
use. Appendix \ref{betafun} contains the details of the
calculation of the beta functions at weak coupling in the
nonlinear sigma model.

\section{Transfer matrices and supersymmetry}
\label{tmatr}

In this Section we will express the Ising model transfer matrices
in terms of fermionic replicas, and then introduce bosons to make
the system supersymmetric. This allows us to consider (in Sec.\
\ref{enhsusy}) averages over quenched disorder without taking the
replica $n \to 0$ limit.

First, we set up some notation. We define the dual coupling
$\tilde K$ by \beq e^{-2\tilde K_{\bi\bj}} = \tanh K_{\bi\bj} \eeq
for any sign of $K_{\bi\bj}$. For positive $K_{\bi\bj} = K > 0$ we
denote $\tilde K_{\bi\bj} = K^*$, and for negative $K_{\bi\bj} =
-K <0$, $\tilde K_{\bi\bj} = K^* + i\pi/2$. We assume free (not
periodic) boundary conditions on the Ising spins in the horizontal
($x$) direction, and periodic in the vertical ($y$) direction. As
is well known (see, for example, Ref. \onlinecite{baxter}), the
partition function of the nearest-neighbor Ising model may be
written as the trace of a product of row transfer matrices: \beq Z
= \Tr \prod_{i_y} T_v(i_y) T_h(i_y). \label{partfun} \eeq Here
$i_y$ is an integer coordinate of a row of sites. The  row
transfer matrices, $T_v(i_y)$ for vertical, and $T_h(i_y)$ for
horizontal bonds, do not commute with each other, so the product
in Eq.\ (\ref{partfun}) must be ordered such that the row
coordinate $i_y$ increases from right to left. The row transfer
matrices may in turn be written as products of the transfer
matrices for single bonds: \beq T_v(i_y) = \prod_{i_x} T_{v\bi},
\qquad T_h(i_y) = \prod_{i_x} T_{h\bi}. \eeq The $T_{v\bi}$'s for
different $i_x$ and the same $i_y$ commute, and similarly for the
$T_{h\bi}$'s. The trace represents the periodic boundary condition
in the $y$ direction.

Following Ref. \onlinecite{sml} we write the vertical transfer matrix for
a single vertical bond between the Ising spins at points $\bi$ and
$\bi + \by$ as
\beq
T_{v\bi} =
\frac{e^{K_{\bi,\bi+\by}}}{\cosh\tilde K_{\bi,\bi+\by}}
\exp\left(\tilde K_{\bi,\bi+\by} \sx_i\right),
\eeq
where the ``light'' index $i$ denotes a site on the 1D lattice
corresponding to the vertical row containing the original site $\bi$, and
$\sx$, $\sy$ and $\sz$ are Pauli matrices. Similarly, the horizontal
transfer matrix for a single horizontal bond between the Ising spins at
points $\bi$ and $\bi + \bx$ is
\beq
T_{h\bi} = \exp\left(K_{\bi,\bi + \bx} \sz_i \sz_{i+1}\right).
\eeq
Note that before the averaging over the randomness the transfer matrices
explicitly depend on the corresponding bond, and therefore are labeled
by the bold 2D indices.

The transfer matrices act in tensor products of two-dimensional
spaces at each horizontal coordinate $i_x$. These 2D spaces may be
realized as Fock spaces of fermions on a 1D chain of sites. This
fermionization is implemented by the Jordan-Wigner transformation
relating Pauli matrices to fermionic operators. To use this
transformation we first make a canonical transformation \beq \sx_i
\to -\sz_i, \qquad \sz_i \to \sx_i. \eeq The Jordan-Wigner
transformation reads \bea \sz_i &=& 2 \cd_i \cph_i -1, \\ \sx_i
\sx_{i+1} &=& (\cd_i - \cph_i)(\cd_{i+1} + \cph_{i+1}), \eea where
$\cd_i$ and $c_i$ are canonical creation and annihilation
fermionic operators. In terms of these operators the transfer
matrices for individual bonds become \bea T_{v\bi} &=&
\frac{e^{K_{\bi,\bi+\by}}}{\cosh\tilde K_{\bi,\bi+\by}}
\exp\left(- 2\tilde K_{\bi,\bi+\by} (\cd_i\cph_i - 1/2)\right),
\label{Tv} \\ T_{h\bi} &=& \exp\left(K_{\bi,\bi + \bx}(\cd_i -
\cph_i)(\cd_{i+1} + \cph_{i+1}) \right). \label{Th} \eea

Next we replicate the fermions. The number of replicas has to be
even, because of the symmetry of the bosons to be introduced
below, so we denote it as $2n$. We label the replicas by Greek
letters. The replicated transfer matrices become \bea T_{v\bi} &=&
(2\cosh K)^{2n} \exp\left(-2 \tilde K_{\bi,\bi+\by} \Nfi \right),
\\ T_{h\bi} &=&  \exp\left(2 K_{\bi,\bi + \bx} \Xfi \right), \eea
where we defined \bea \Nfi &=& \sum_{\al = 1}^{2n} n_{\al i},
\qquad n_{\al i} = \cd_{\al i} \cph_{\al i}, \\ \Xfi &=& \sum_{\al
= 1}^{2n} x_{\al i}, \\ x_{\al i} &=& \frac{1}{2}(\cd_{\al i} -
\cph_{\al i}) (\cd_{\al, i+1} + \cph_{\al, i+1}). \label{x} \eea

The quadratic forms $\Nfi$ and $\Xfi$ are invariant under the
orthogonal transformations mixing the fermions, which becomes
especially transparent if we introduce two sets of $2n$ real
fermions per site as \beq \eta_{\al i} = {\cd_{\al i} - \cph_{\al
i} \over \sqrt{2} i}, \qquad \xi_{\al i} = {\cd_{\al i} +
\cph_{\al i} \over \sqrt{2}}. \eeq These fermions satisfy \beq
\{\eta_{\al i}, \eta_{\be j}\} = \{\xi_{\al i}, \xi_{\be j}\} =
\delta_{ij} \dab, \quad \{\eta_{\al i}, \xi_{\be j}\} = 0.
\label{anticomm} \eeq Terminologically, we note here that any set
of self-adjoint operators, say $\psi_a$, $a=1$, \ldots $M$, for
some $M$, with anticommutation relations
$\{\psi_a,\psi_b\}=\delta_{ab}$, constitutes a {\em Clifford
algebra}. For us, the set of $\xi$'s, either for one or for many
sites, or similarly of $\eta$'s, or a combination of these, are
all Clifford algebras. A little of the general theory of these
algebras will be used later. In terms of these fermions, or
Clifford algebra generators, the quadratic forms become \beq \Nfi
= i \eta_{\al i} \xi_{\al i} + n, \qquad \Xfi = i\eta_{\al i}
\xi_{\al, i+1}, \eeq where from now on we assume that repeated
indices from the beginning of the Greek alphabet ($\al$, $\be$,
etc.) are summed from 1 to $2n$, unless stated otherwise.

The generators of the global symmetry algebra so($2n$), in this
notation, are $\sum_i(\eta_{\al i} \eta_{\be i} + \xi_{\al i}
\xi_{\be i})$, for pairs $\alpha$, $\beta$, and because of the
anticommutation relations we may take only $\alpha<\beta$,
corresponding to the antisymmetric $2n\times 2n$ matrices. These
generators commute with $\Nfi$ and $\Xfi$, proving that the
transfer matrices are invariant under so($2n$). The replicated
partition function $Z^{2n}$, which is now given by a trace in the
$2n$-component fermion Fock space, is invariant under so($2n$).
Note that we capitalize the name of the group or supergroup, such
as SO($2n$), but not the name of the corresponding Lie
(super-)algebra, such as so($2n$).

The supersymmetric counterpart of the fermionic algebra so$(2n)$
is the symplectic algebra sp$(2n)$. This motivates the
introduction of bosons with this symplectic symmetry as follows.
We start with $2n$ complex ``symplectic'' bosonic operators
$s_{\al i}$ (and their adjoints $\sd_{\al i}$), satisfying \beq
[\sph_{\al i}, \sd_{\be j}] = i \de_{ij} \Jab, \eeq where $\jab$
is a non-singular real antisymmetric $2n \times 2n$ matrix. (It is
because the number of bosons must be even that the number of
fermions must be also.) Without loss of generality (by appropriate
change of basis) this matrix may be taken to be (in block form)
\beq J = \left(
\begin{array}{cc}
        0 & \openone_n \\
        -\openone_n & 0
        \end{array}
\right), \eeq where $\openone_n$ is the $n\times n$ identity
matrix. We also need to define the vacuum state for our bosons. We
will see that the SUSY requirement makes this choice essentially
unique, and the resulting space of states has indefinite metric
(some states have negative squared norms).

The bosonic counterparts of the forms $\Nfi$ and $\Xfi$ are the
symplectic forms \bea \Nbi &=& i \, \sd_{\al i} \Jab \sph_{\be i},
\\ \Xbi &=& \frac{i}{2} (\sd_{\al i} - \sph_{\al i}) \Jab
(\sd_{\be, i+1} + \sph_{\be, i+1}). \label{bosons} \eea To
parallel the fermionic case we also introduce two sets of $2n$
real bosons per site as \beq q_{\al i} = {\sd_{\al i} + \sph_{\al
i} \over \sqrt{2}}, \qquad r_{\al i} = {\sd_{\al i} - \sph_{\al i}
\over \sqrt{2} i}. \eeq These bosons satisfy \beq [q_{\al i},
q_{\be j}] = [r_{\al i}, r_{\be j}] = i \de_{ij} \jab, \qquad
[q_{\al i}, r_{\be j}] = 0. \label{commrel} \eeq (These have the
form of the commutation relations for canonically conjugate
coordinates and momenta.) In terms of the real bosons the forms
(\ref{bosons}) are \beq \Nbi = - r_{\al i} \jab q_{\be i} - n,
\quad \Xbi = - r_{\al i} \jab q_{\be, i+1}. \eeq

The generators of the global symplectic symmetry algebra sp($2n$)
are $\sum_i(q_{\al i} q_{\be i} + r_{\al i} r_{\be i})$, where
because of the commutation relations we may use only
$\alpha\leq\beta$, corresponding to symmetric matrices. These
operators are the generators of global linear canonical
transformations on the $q$'s and $r$'s. The forms $\Nbi$ and
$\Xbi$, and hence the transfer matrices, are invariant under this
algebra.

We now address the question of the bosonic vacuum. We will find it
using the requirement that the spectrum of the bosonic form $\Nbi$
is the SUSY counterpart of the integer spectrum of the fermionic
form $\Nfi$. In this case we will have \beq \STr \exp\left(-{\mbox
{\rm const}}(\Nfi +\Nbi)\right) = 1. \label{susycond} \eeq This
condition is essential in the SUSY approach to ensure that the
partition function of the RBIM is unity for any realization of the
disorder. Here we used the notation ${\rm STr}$ for the supertrace
in the space of states of our problem. We will now discuss how
this supertrace is defined.

In general, the supertrace in a super-vector space must be defined
using the notion of a grading for the states (or vectors). This
can be done by choosing a basis and then defining one subset of
basis vectors as ``even'', and the remainder as ``odd'', vectors.
The vector space then contains two complementary subspaces of even
and odd vectors, respectively; the zero vector, and linear
combinations of vectors from both subspaces, are viewed as having
no definite grading. The vectors are then said to be ${\Bbb
Z}_2$-graded. Operators on the vector space can likewise be
classified as even or odd, according to whether they preserve or
reverse the grading of basis vectors on which they act; usually,
only operators for which this rule gives a consistent answer
(those with well-defined grading) are of interest. Thus the
grading is usually treated as a superselection rule. The
supertrace $\STr Y$ of an even operator $Y$ is then defined, like
an ordinary trace, as the sum of the diagonal matrix elements of
$Y$ in a basis of even and odd vectors, except that for the
supertrace, the diagonal elements in the odd basis vectors are
weighted by a minus sign. (Note that the matrix elements $Y_{IJ}$
of an operator $Y$ are obtained as the coefficients in the system
of equations $Y|J\rangle=\sum_IY_{IJ}|I\rangle$, where
$|I\rangle$, $I=1$, $2$, \ldots, are the basis vectors, {\em
without} using an inner product on the vector space.) The
supertrace has a number of nice properties, like the ordinary
trace in an ordinary vector space; in particular a form of the
cyclic property still holds, $\STr AB=\pm \STr BA$, with a $+$ if
both $A$ and $B$ are even, and a $-$ if both are odd operators.
The grading and the supertrace are needed in connection with
supersymmetry algebras, but otherwise do not necessarily have to
be considered. We also note here that it is possible to form a
graded tensor product of graded vector spaces, in a way that
preserves the grading.

In this paper, most of our constructions use a Fock space. In a
Fock space generated by boson and fermion operators acting on a
vacuum, there is a natural grading, defined using an occupation
number basis, in which states are even or odd according as the
total number of fermions (of all types) is even or odd. However,
we will {\em not} use such a grading to define the supertrace
above. The reason is that we have already introduced the ordinary
trace in writing the partition function for fermionic replicas; in
this trace, all diagonal matrix elements are taken with weight
$+1$, including those in states with an odd number of fermions. It
is of course quite standard to use an ordinary trace even when
dealing with fermions, which have a natural grading. The natural
grading is used in defining a tensor product, such that fermion
operators on different sites (i.e.\ in different factors in the
tensor product) anticommute. These are the tensor products usually
used by physicists for second quantized fermion problems. Each
time we write a tensor product of spaces, it will be the graded
tensor product using the natural grading that we mean. There is
nothing wrong with the use of the trace, unless we are concerned
about SUSY. The grading that we use in introducing SUSY into our
representation is defined by specifying that states with an even
(odd) number of {\em bosons} are even (odd), and so \beq \STr
\ldots = {\rm Tr}(-1)^{\sum_i \Nbi} \ldots. \eeq For states with
no bosons, this reduces to the usual trace.

Now that we have defined the supertrace, we must arrange to
satisfy the condition (\ref{susycond}). The form $\Nbi$ may be
diagonalized with the transformation to two other sets of $n$
complex bosons. Namely, we define \beq a_{\mu i} = {s_{\mu i} + i
s_{\mu + n, i} \over \sqrt{2}}, \qquad \ab_{\mu i} = {s_{\mu i} -
i s_{\mu + n, i} \over \sqrt{2}}, \label{defaab} \eeq and the
adjoint operators, where the index $\mu$ (and other indices from
the middle of the Greek alphabet, like $\nu$, etc.) runs from 1 to
$n$. These bosons satisfy \beq [\aph_{\mu i}, \ad_{\nu j}] =
\de_{ij} \de_{\mu \nu}, \qquad [\abph_{\mu i}, \abd_{\nu j}] = -
\de_{ij} \de_{\mu \nu}, \eeq and the rest of commutators vanish.
In terms of these bosons we have \beq \Nbi = \ad_{\mu i} \aph_{\mu
i} - \abd_{\mu i} \abph_{\mu i}. \label{diagonal_form} \eeq If we
introduce the vacuum for $a$ and $\ab$ bosons in the usual manner
\beq a_{\mu i} |0\ra =\ab_{\mu i} |0\ra = 0, \label{defvac} \eeq
then the spectrum of $\Nbi$ is the non-negative integers, which is
the SUSY counterpart of the spectrum of $\Nfi$. This ensures that
Eq.\ (\ref{susycond}) holds. But the price to pay is that the
states with an odd number of $\ab$ bosons have negative norms. By
another choice of the vacuum we could avoid negative norms, but
then we would not have the supersymmetry. With these definitions,
we have now defined a Fock space $\cal F$, which is a tensor
product of Fock spaces at each site, ${\cal F}=\otimes_i{\cal
F}_i$ in an obvious notation. The tensor product of Fock spaces is
defined using the natural grading, however our choice of grading
also behaves well in the product; the grading of states is
determined by the product of the ``degrees'' ($=\pm 1$ for even,
odd respectively) of the states on the sites, because boson
numbers add. Note that fermion operators are viewed as even in our
grading.

The transfer matrices including fermions and bosons
supersymmetrically are now \bea T_{v\bi} &=& \exp\left(-2\tilde
K_{\bi,\bi+\by} \NSi \right), \no\\ T_{h\bi} &=& \exp\left(2
K_{\bi,\bi + \bx} \XSi \right), \label{tvth} \eea where the
subscript $S$ stands for ``supersymmetric'', \beq \NSi = \Nfi +
\Nbi, \qquad \XSi = \Xfi + \Xbi. \eeq The SUSY transfer matrices
are invariant under the {\it orthosymplectic} superalgebra {\os},
since the forms $\NSi$ and $\XSi$ commute with the generators
$\sum_i(\xi_{\al i} \xi_{\be i} + \eta_{\al i} \eta_{\be i})$,
$\sum_i(q_{\al i} q_{\be i} + r_{\al i} r_{\be i})$, and
$\sum_i(\xi_{\al i} q_{\be i} + \eta_{\al i} r_{\be i})$ of {\os}.
The last set of generators are the ``odd'' (with respect to either
grading), fermionic, or supergenerators of the superalgebra, and
$\alpha$ and $\beta$ can take arbitrary values there. Note that,
in a superalgebra, two even operators obey commutation relations,
two odd operators anticommutation relations, and an even with an
odd generator obeys a commutation relation. Thus the definition of
the superalgebra structure again involves the grading. The
definition of the supertrace also respects supersymmetry.

The condition (\ref{susycond}) applies when only vertical-bond
transfer matrices are present. To prove the supersymmetry of the
full problem, namely, that the supersymmetrized partition function
$Z_{\rm SUSY}$ (the supertrace of the product of supersymmetrized
transfer matrices) is unity for any realization of the disorder,
we use a graphical representation of $Z$. Imagine a
high-temperature expansion of the horizontal transfer matrix
$T_{h\bi}$, where we expand it in powers of $K_{\bi,\bi + \bx}$.
At each horizontal row the operator \bea 2\XSi &=& (\cd_{\al i} -
\cph_{\al i}) (\cd_{\al, i+1} + \cph_{\al, i+1}) \br + i (\sd_{\al
i} - \sph_{\al i}) \Jab (\sd_{\be, i+1} + \sph_{\be, i+1}) \eea
may create or destroy two particles at neighboring sites, or it
may transfer one particle between neighboring sites. The vertical
transfer matrices are diagonal, so they only propagate particles
in the vertical direction.

We can represent these processes graphically by lines starting and
ending at lattice sites (creation and annihilation), joining the
horizontal pairs of sites (hopping), and joining vertical pairs of
site (propagation due to $T_v$). When two such lines start at a
pair of neighboring sites on a row and end at another pair of
neighboring sites on a row, we call this a closed loop (similar to
closed loops in high temperature expansion of the pure Ising
model). Then the supersymmetrized partition function is equal to 1
plus the sum of contributions of all closed loops. Then we have to
prove that for a closed loop the contributions of fermions and
bosons cancel each other.

Let us take the smallest possible loop, where two particles are
created and destroyed on two adjacent rows. This is represented by
two short vertical lines between two neighboring pairs of lattice
sites. For a given fermionic replica, say 1, this loop contributes
the following term: \bea &&-\la 0|\cph_{1i}\cph_{1,i+1}e^{-2\tilde
K_{\bi,\bi+\by} \Nfi} \br \times e^{-2\tilde
K_{\bi+\bx,\bi+\bx+\by} {\hat N}_{F,i+1}} \cd_{1i}\cd_{1,i+1}
|0\ra = \br e^{-2(\tilde K_{\bi,\bi+\by} + \tilde
K_{\bi+\bx,\bi+\bx+\by})}. \label{fermiloop} \eea The
corresponding bosonic contribution is \bea &&\la 0|i\sph_{n+1,i}
\sph_{1,i+1} e^{-2\tilde K_{\bi,\bi+\by} \Nbi} \br \times
e^{-2\tilde K_{\bi+\bx,\bi+\bx+\by} {\hat N}_{B,i+1}} i \sd_{1i}
\sd_{n+1,i+1} |0\ra. \label{boseloop} \eea To evaluate this
expression we note the following. {}From the definition of the
symplectic bosons it follows that \beq [\sph_{\al j}, \Nbi] =
\de_{ij} \sph_{\al j}, \qquad [\sd_{\al j}, \Nbi] = - \de_{ij}
\sd_{\al j}. \eeq We use these relations to pull the exponentials
through to the vacuum on the right in Eq.\ (\ref{boseloop}), which
becomes then \beq -\la 0|\sph_{n+1,i} \sph_{1,i+1} \sd_{1i}
\sd_{n+1,i+1} |0\ra e^{-2(\tilde K_{\bi,\bi+\by} + \tilde
K_{\bi+\bx,\bi+\bx+\by})}. \eeq Next we notice that, as a
consequence of Eq.\ (\ref{defaab}) and the definition
(\ref{defvac}), the vacuum state $|0\ra$ is annihilated by
$\sph_{\al i}$. Then we commute the operators in the first factor
in the last expression, after which it becomes exactly opposite to
the fermionic contribution (\ref{fermiloop}). This argument is
easily generalized to arbitrary loops (including those that wrap
around the system, thanks to the definition of the supertrace),
and proves that the supersymmetrized partition function is indeed
equal to one, for any realization of the disorder.

\section{Averaging and enhanced supersymmetry on the Nishimori line}
\label{enhsusy}

In this Section we perform the average with respect to the
distribution $P[K]$ and find that on the NL the averaged transfer
matrices have {\it enhanced} supersymmetry osp($2n+1\,|\,2n$). In
Appendix \ref{eqcorr} we use this enhanced SUSY to rederive the
equality (\ref{correq}) for the Ising correlators.

Here we need some more notation. We introduce a parameter of the
form of the Ising coupling, $L$, and its dual $L^*$, related to
the probability $p$: \beq 1-2p = \tanh L = e^{-2 L^*}. \eeq In
terms of $L$ the equation (\ref{nishimori}) of the NL is $L = K$.
Below the NL $L < K$, and above the NL $L > K$.

Since the couplings $K_{\bi\bj}$ are independent, we can average
transfer matrices for different bonds separately. For a vertical
transfer matrix this gives (recall that the disorder averages are
denoted by square brackets)
\beq T_{1i} = \disave{T_{v\bi}} =
\exp\left(-2 K^* \NSi\right) \left(1 - p + p(-1)^{\NSi}\right).
\label{t1} \eeq
Note that after the averaging the translational
invariance is restored, and this allows us to label the average
transfer matrices by 1D (``light'') indices.

The value of the last factor in Eq.\ (\ref{t1}) depends on the
value of $\NSi$ in the state, on which $T_{1i}$ acts. For an even
$\NSi$ it equals 1, for an odd $\NSi$ it gives $1-2p = e^{-2L^*}$.
Then we can rewrite the operator (\ref{t1}) in a slightly
different form. Namely, we introduce additional (zeroth) fermionic
state and operators $\cd_{0i}$, $\cph_{0i}$ and consider the
subspace ${\cal F}'_i$ given by the following constraint: \beq
{\hat N}'_{Si} = \NSi + n_{0i} = \mbox{\rm \ even}, \label{space}
\eeq where $n_{0i}$ is defined in analogy with $n_{\al i}$: \beq
n_{0i} = \cd_{0i} \cph_{0i}. \eeq That is, the number of fermions
plus bosons on each site must be even. There is a one-to-one
correspondence between the states in this subspace and the
original Fock space ${\cal F}_i$. This correspondence is
illustrated for the case $n=1$ in the Table \ref{tab1}. The
grading in the space ${\cal F}'=\otimes_i {\cal F}'_i$ is taken to
be the same as that in ${\cal F}$, which was not the natural
grading. However, we see that in a Fock space with a constraint of
the form of Eq.\ (\ref{space}), the number of bosons is odd if and
only if the number of fermions is odd (this is true for each site
and also for the tensor product). Hence, our grading on ${\cal
F}'$ is the same as the one obtained from the natural grading on
the larger Fock space with $2n+1$ fermion species, when restricted
to the subspace. Now we can replace $T_{1i}$ by the operator \beq
T'_{1i} = \exp\left(-2 K^* \NSi - 2 L^* n_{0i}\right), \label{t1c}
\eeq which has the same matrix elements in the constrained
subspace ${\cal F}'_i$, as $T_{1i}$ had in the original space
${\cal F}_i$ (between the corresponding states). {}From now on in
this Section we will denote transfer operators acting in the
constrained spaces ${\cal F}'$ by a prime.

\begin{table}
\begin{center}
\begin{tabular}{|c|c|}
\makebox[42mm]{${\cal F}_i$} &
\makebox[41mm]{${\cal F}'_i$} \\
\hline
$|0\ra$ & $|0\ra$ \\
$\cd_{1i}|0\ra$ & $\cd_{0i}\cd_{1i}|0\ra$ \\
$\cd_{2i}|0\ra$ & $\cd_{0i}\cd_{2i}|0\ra$ \\
$\cd_{1i}\cd_{2i}|0\ra$ & $\cd_{1i}\cd_{2i}|0\ra$ \\
$\ad_{1i}|0\ra$ & $\cd_{0i}\ad_{1i}|0\ra$ \\
$\cd_{1i}\ad_{1i}|0\ra$ & $\cd_{1i}\ad_{1i}|0\ra$ \\
$\cd_{2i}\ad_{1i}|0\ra$ & $\cd_{2i}\ad_{1i}|0\ra$ \\
$\cd_{1i}\cd_{2i}\ad_{1i}|0\ra$ &
$\cd_{0i}\cd_{1i}\cd_{2i}\ad_{1i}|0\ra$ \\
\vdots & \vdots \\
\end{tabular}
\end{center}
\vspace{2mm}
\caption{Correspondence between states in the spaces ${\cal F}_i$
and ${\cal F}'_i$.}
\label{tab1}
\end{table}

For a horizontal transfer matrix the averaging gives \beq T_{2i} =
\disave{T_{h\bi}} ={\cosh(2K\XSi + L) \over \cosh L}. \label{t2}
\eeq To find the corresponding operator in ${\cal F}'$ we need to
establish some substitution rules for basic operators.

Single creation and annihilation operators like $\cd_{1i}$, which
are quite legitimate in the space ${\cal F}_i$, do not act within
${\cal F}'_i$. Using the correspondence between the states, given
in Table \ref{tab1}, it is easy to establish, that in ${\cal
F}'_i$ the operator $\cd_{1i}$ must be replaced by the operator
$(\cd_{0i} - \cph_{0i})\cd_{1i}$. However, this operator is
bosonic (i.e.\ even with respect to the natural grading on ${\cal
F}'$), so in the tensor product ${\cal F}' = \otimes_i {\cal
F}'_i$ it will not have the anticommutation properties that
$\cd_{1i}$ had in $\cal F$. To make it fermionic in the total
space ${\cal F}'$, we need to attach to it a string: \bea \cd_{1i}
& \to & (\cd_{0i} - \cph_{0i})\cd_{1i} \Sigma_{0i},
\\ \Sigma_{0i} & = & \prod_{j > i}(-1)^{n_{0j}}. \label{cinf'}
\eea Repeating this argument for other operators, we obtain the
following rules of substitution: \bea \cd_{\al i} &\to& (\cd_{0i}
- \cph_{0i})\cd_{\al i} \Sigma_{0i},\no\\ \cph_{\al i} &\to&
(\cd_{0i} - \cph_{0i})\cph_{\al i} \Sigma_{0i},\no\\ \ad_{\mu i}
&\to& (\cd_{0i} + \cph_{0i})\ad_{\mu i} \Sigma_{0i},\no\\
\aph_{\mu i} &\to& (\cd_{0i} + \cph_{0i})\aph_{\mu i}
\Sigma_{0i},\no\\ \abd_{\mu i} &\to& (\cd_{0i} +
\cph_{0i})\abd_{\mu i} \Sigma_{0i},\no\\ \abph_{\mu i} &\to&
(\cd_{0i} + \cph_{0i})\abph_{\mu i} \Sigma_{0i}. \label{subst}
\eea As an alternative to looking at all the states, the
correspondence can be established by verifying that the right-hand
sides of these expressions have the same (anti-)commutators as the
left-hand sides. It follows from these rules that any product of
an even number of creation and/or annihilation operators in ${\cal
F}_i$ remains the same, when going to ${\cal F}'_i$.

Now note what happens with $x_{\al i}$ (see Eq.\ (\ref{x})) upon
transition from ${\cal F} = \otimes_i {\cal F}_i$ to ${\cal F}'$:
\bea 2 x_{\al i} &=& (\cd_{\al i} - \cph_{\al i}) (\cd_{\al, i+1}
+ \cph_{\al, i+1}) \no\\ &\to& (\cd_{0i} - \cph_{0i})(\cd_{\al i}
- \cph_{\al i}) (-1)^{n_{0,i+1}} \br \times (\cd_{0,i+1} -
\cph_{0,i+1}) (\cd_{\al, i+1} + \cph_{\al, i+1}) \no\\ &=& 4
x_{0i} x_{\al i}, \label{xxx} \eea where $x_{0i}$ is defined
similarly to Eq.\ (\ref{x}): \beq x_{0i} = \frac{1}{2}(\cd_{0i} -
\cph_{0i})(\cd_{0, i+1} + \cph_{0, i+1}) = i \eta_{0i}
\xi_{0,i+1}. \eeq

With these substitution rules established, we can see that in the
space ${\cal F}'$ the operator corresponding to $T_{2i}$ is given
by \beq T'_{2i} = {\cosh(2 K \XSi + 2 L x_{0i}) \over \cosh L},
\label{t2c} \eeq which easily follows from the substitution rule
(\ref{xxx}) and the fact that $x_{\al i}^2 = x_{0i}^2 = 1/4$.
{}From Eqs.\ (\ref{t1c}), (\ref{t2c}), we see that the transfer
matrices commute with the constraint, Eq.\ (\ref{space}). In the
full space of states that includes states of odd, as well as even,
fermion plus boson number at each site, there are local ${\Bbb
Z}_2$ operations given by $(-1)^{{\hat N}'_{Si}}$. This is
therefore a gauge symmetry under which the allowed states and
transfer operators (and also physical observables) must be
invariant.

The forms (\ref{t1c}), (\ref{t2c}) are very convenient for the
discussion of the symmetry properties of our model on the NL.
Indeed, we see that on the NL, where $K = L$, the operators $T'_i$
become \bea T'_{1i} &=& \exp\left(-2 K^* {\hat N}'_{Si} \right),
\label{t1n} \\ T'_{2i} &=& {\cosh(2 K \XSi') \over \cosh K},
\label{t2n} \eea where \beq \XSi' = \XSi + x_{0i} = i \eta_{ai}
\xi_{a, i+1} - r_{\al i} \jab q_{\be, i+1}, \eeq and the Latin
subscripts from the beginning of the alphabet denote the fermionic
indices running form 0 to $2n$.

On the NL, the expressions (\ref{t1n}), (\ref{t2n}) have enhanced
supersymmetry: they are now invariant under an {\osp} algebra. The
generators of this algebra have a similar form as before, but
involve the $2n+1$ fermion operators: $\sum_i(\xi_{a i} \xi_{b i}
+ \eta_{a i} \eta_{b i})$, $\sum_i(q_{\al i} q_{\be i} + r_{\al i}
r_{\be i})$, and $\sum_i(\xi_{a i} q_{\be i} + \eta_{a i} r_{\be
i})$. The last set of generators are the odd ones, with respect to
our grading, or to the natural one on the Fock space of which
${\cal F}'$ is a subspace; we have seen these are equivalent in
the constrained subspace.

As anticipated in the Introduction, this enhanced {\it continuous}
SUSY replaces the gauge symmetry of Nishimori, and the enhanced
permutational symmetry of the replica approach, previously known
to exist in Ising spin language on the NL \cite{n,ldgh}. The
symmetry has many consequences, such as the equalities
(\ref{correq}) among different correlation functions on the NL. In
appendix \ref{eqcorr}, we briefly show how these equalities may be
obtained from the enhanced SUSY exhibited in this section. We have
also obtained the local ${\Bbb Z}_2$ gauge symmetry anticipated in
the Introduction.

\section{Structure of the space of states and the Hamiltonian limit}
\label{hamlimit}

In this Section we first analyze (in Sec.\ \ref{supspin}) the
structure of the space of states of our quantum problem and then
take the time continuum limit of our transfer matrices and obtain
a quantum Hamiltonian describing our system. This has the form of
a spin chain with {\em irreducible} representations of the
symmetry algebra {\osp} at each site. Then in Sec.\ \ref{unconstr}
we consider a more explicit construction of these irreducible
representations.

\subsection{Superspin chain and Hamiltonian limit} \label{supspin}

Let us consider the structure of the constrained space ${\cal
F}'_i$ (\ref{space}), with its natural grading, under
transformations of {\osp}. It is easy to see that ${\cal F}'_i$ is
not irreducible under this algebra. Rather, it has the structure
of the tensor product of two irreducible spinors of {\osp}.

Indeed, let us consider first the fermionic replicas only, i.e.
the replica approach where $n \to 0$ in the end. Then we have the
modified transfer matrices which are invariant under the
orthogonal algebra so($2n+1$), and the subspace they act on is
given at each site by the constraint \beq \Nfi' = \Nfi + n_{0i} =
\mbox{\rm even}. \label{constr} \eeq This space has dimension
$2^{2n}$, and, under so($2n+1$), it transforms as the tensor
product of two spinors of so($2n+1$), each of dimension $2^n$.
These two spinors can be identified as the spaces on which the two
parts $\xi_{ai}\xi_{bi}$, $\eta_{ai}\eta_{bi}$ of the generators
$\xi_{ai}\xi_{bi}+\eta_{ai}\eta_{bi}$ act. The tensor product
decomposes into irreducible representations of so($2n+1$)
corresponding to each even value of $\Nfi'$ in the range 0 to $2n$
allowed by the constraint (\ref{constr}). Similarly, the
orthogonal subspace $\Nfi'=\mbox{\rm odd}$ is also a tensor
product of spinors and has a similar decomposition.

When the bosons are included as in the SUSY approach, the two
parts $q_{\alpha i}q_{\beta i}$, $r_{\alpha i}r_{\beta i}$ of the
sp($2n$) generators at a site $i$ generate infinite-dimensional
spinor representations of sp($2n$) (sometimes known as metaplectic
representations). When the fermions and bosons are combined
together with the constraint that ${\hat N}_{Si}'$ be even, the
resulting space is a tensor product of irreducible spinors of
{\osp}. The fact that a single such tensor product is involved is
the nontrivial part of this statement, and is addressed further in
Sec.\ \ref{unconstr}. These spinors comprise one lowest-weight
representation of {\osp}, which we denote by $R$, and one highest
weight representation of {\osp}, which we denote by $\bar R$. Thus
we may write ${\cal F}_i'=R_i\otimes \bar{R}_i$.

This organization of states suggests a picture of our model as a system of
``superspins'' (spinors $R$ and $\bar R$ of {\osp}) sitting in pairs on
the sites of the 1D lattice. It is convenient to combine the corresponding
generators of {\osp} into square matrices consistent with reality
properties satisfied by the matrices of {\osp} in the defining
representation. Namely, the generators of {\osp} acting in the
representation $R$ are combined into the superspin
\beq
G = \left( \begin{array}{cc}
        \xi_{a} \xi_{b} - {1 \over 2} \de_{ab} &
        \xi_{a} q_{\be} \\
        i J_{\al \ga} q_{\ga} \xi_{b} &
        i J_{\al \ga} q_{\ga} q_{\be} - {1 \over 2} \dab
        \end{array} \right),
\label{suspin}
 \eeq
shown here in block $\bigl((2n+1)+2n\bigr) \times
\bigl((2n+1)+2n\bigr)$ form, and a similar matrix obtained from
the generators of {\osp} acting in the representation $\bar R$:
\beq {\bar G} = \left(
\begin{array}{cc}
        \eta_{a} \eta_{b} - {1 \over 2} \de_{ab} &
        \eta_{a} r_{\be} \\
        i J_{\al \ga} r_{\ga} \eta_{b} &
        i J_{\al \ga} r_{\ga} r_{\be} - {1 \over 2} \dab
        \end{array} \right).
\label{suspinb} \eeq The generators of the global SUSY of the
system are now, in matrix form, $\sum_i(G_i + \bar{G}_i)$.

Next we take a time-continuum (Hamiltonian) limit of the transfer
matrices. To do that we actually have to start with an anisotropic
RBIM, where the vertical and horizontal couplings take different
values of $K$. In that case we can arrange for the situation when
in the vertical matrix $K^*, L^* \ll 1$, and in the horizontal one
$K, L \ll 1$ (we continue to use these notations, so that $K^*$,
$L^*$ are no longer related to $K$, $L$ by the duality relation).
Then we can expand the horizontal transfer matrix (\ref{t2c}) as
\bea T'_{2i} &\simeq& 1 + 2 K^2 \XSi^2 + 4 K L \XSi x_{0i} \no\\
&\simeq& \exp\left(2 K^2 \XSi^2 + 4 K L \XSi x_{0i}\right),
\label{t'2cont} \eea Then we can combine all $T'_{1i}$ and
$T'_{2i}$ into a single ``evolution'' operator in imaginary time
with the Hamiltonian \beq H_S = \sum_i \left(h \bigl({\hat N}_{Si}
+ \lam^* n_{0i}\bigr) - k \bigl(\XSi^2 + 2\lam\XSi x_{0i}
\bigr)\right), \label{ham1} \eeq where we introduced \beq h = 2
K^*, \quad k = 2 K^2, \quad \lam^* = \frac{L^*}{K^*}, \quad \lam =
\frac{L}{K}. \eeq The parameters $\lam$ and $\lam^*$ introduce
anisotropy in the couplings among the replicas.

\begin{figure}
\epsfxsize=3.4in \centerline{\epsffile{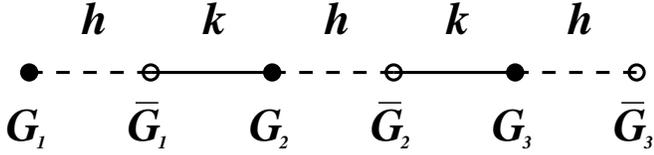}}
\vspace{0.25in} \caption{The graphical representation of the
Hamiltonian (\protect\ref{ham2}) on the {\it split\/} sites.
Superspins in the representations $R$ and $\bar R$ are shown as
filled and empty circles. The two types of coupling, $h$ and $k$,
are indicated.} \label{fig2}
\end{figure}

The Hamiltonian (\ref{ham1}) may be rewritten in a very suggestive
form using the superspins $G$ and ${\bar G}$: \beq H_{S} = \sum_i
\left(h \bigl(\NSi + \lam^* n_{0i}\bigr) + k \str \La {\bar G}_i
\La G_{i+1} \right), \label{ham2} \eeq where \beq \La = \mbox{\rm
diag}(\lam, \openone_{4n}) \eeq is a diagonal matrix representing
the anisotropy in the superspin space. The supertrace here,
denoted ${\rm str}$, is over the $4n+1$-dimensional space as
above, with $+$ for diagonal matrix elements in the
$2n+1$-dimensional block, and $-$ for those in the remaining
$2n$-dimensional block. With this definition, the expressions
reduce to the ordinary trace for the replica formalism where only
the $\xi$'s and $\eta$'s are kept (and then $n \to 0$), without an
overall change in sign.

Now we should notice that in the anisotropic version of the RBIM,
there are in general two couplings, $K_x$, $K_y$, and two
parameters for the probabilities, $L_x$, $L_y$. The Nishimori
condition becomes two equations, $K_x=L_x$, $K_y=L_y$. Thus the NL
is replaced by a two-dimensional surface (or 2-surface) in the
four-dimensional space, and so does not divide the phase diagram
into two pieces. We will continue to refer to this as the NL.
There is presumably a line on this surface at which a transition
occurs. The complete phase boundary is three-dimensional, and the
multicritical behavior is found on a 2-surface on this 3-surface.
The multicritical line on the NL presumably lies in the
multicritical 2-surface on the phase boundary. Even though the
transfer matrices do not have the larger SUSY everywhere on that
2-surface, we presume by universality that the higher SUSY fixed
point theory, to which the multicritical point on the NL flows,
controls the entire multicritical 2-surface, because anisotropy
such as we have introduced usually does not affect the
universality class.

In any case, on the NL $\lam = \lam^* = 1$, and we obtain an
{\osp}-invariant (or isotropic) Hamiltonian \beq H_{S} = \sum_i
\left(h \NSi' + k \str {\bar G}_i G_{i+1} \right). \label{ham3}
\eeq The Eq.\ (\ref{ham3}) has the form of a superspin chain with
the alternating lowest- and highest-weight representations $R$ and
$\bar R$, and corresponding superspin operators $G$ and ${\bar
G}$. We can better represent this by splitting the original sites
into pairs of {\it split\/} sites. This is shown in Fig.\
\ref{fig2}. The Hamiltonian (\ref{ham3}) has two types of
couplings on the alternating bonds. Both these couplings are {\it
antiferromagnetic} in nature. This means that the lowest energy
state for a given bond is the singlet of {\osp} contained in the
decomposition of the tensor product of the representations $R$ and
$\bar R$.

\subsection{Unconstrained representation of superspins}
\label{unconstr}

The picture of the split sites carrying irreducible
representations is very attractive, but does suffer from one
difficulty at present. This is that we obtained the
representations by introducing an additional zeroth fermion
$c_{0i}$, $c_{0i}^\dagger$, together with a constraint which
refers to both the split sites that comprise the original site.
Here and in Appendix \ref{reposp} we show how to avoid this by use
of a different construction. In Sec.\ \ref{general} we will extend
this approach further, introducing a further representation which
involves a constraint on each {\em split} site.

The representations $R$ and $\bar R$ can be constructed using
complex fermionic and bosonic operators without constraints, that
is essentially in ${\cal F}_i$. For the simplest case of the
osp$(3|2)$ algebra this is done in Appendix \ref{reposp}. Here we
note that the complex fermions and bosons used in the construction
are related to the real ones (apart from the zeroth fermion) on
the split sites introduced so far in the following manner: \bea
f_{\mu i} &=& \frac{\xi_{\mu i} + i \xi_{\mu + n, i}}{\sqrt 2},
\qquad \fb_{\mu i} = \frac{i \eta_{\mu i} + \eta_{\mu + n,
i}}{\sqrt 2}, \label{ffermions}\\ b_{\mu i} &=& \frac{q_{\mu i} +
i q_{\mu + n, i}}{\sqrt 2}, \qquad \bb_{\mu i} = \frac{i r_{\mu i}
+ r_{\mu + n, i}}{\sqrt 2}. \label{bbosons} \eea All of these
operators are canonical, except for the $\bar b$ bosons, which are
``negative norm'': \beq [\bbph_{\mu i}, \bbd_{\nu j}] = - \de_{ij}
\de_{\mu \nu}. \eeq In terms of these complex bosons and fermions
the quadratic forms appearing in the transfer matrices
(\ref{tvth}) look especially uniform: \bea \XSi &=& \fd_{\mu, i+1}
\fbd_{\mu i} + \fbph_{\mu i} \fph_{\mu, i+1} + \bd_{\mu, i+1}
\bbd_{\mu i} + \bbph_{\mu i} \bph_{\mu, i+1}, \no\\ \NSi &=&
\fd_{\mu i} \fbd_{\mu i} + \fbph_{\mu i} \fph_{\mu i} + \bd_{\mu
i} \bbd_{\mu i} + \bbph_{\mu i} \bph_{\mu i}. \label{SUSYforms}
\eea In this form the subalgebra {\os} on a split site is
generated by bilinears as before, now of the form $f^2$,
$(f^\dagger)^2$, $f^\dagger f$, $b^2$, \ldots, $f^\dagger b$,
\ldots, for $R$. We saw earlier that the expressions for $\XSi$
and $\NSi$ are invariant under this {\os} algebra. However, the
extension to {\osp} is modified since we do not use the zeroth
fermion $c$. Instead, the additional generators include string
operators such as $(-1)^{n_f}$, see App.\ \ref{reposp}, where
expressions for the generators of {\osp} in the case $n=1$ are
given. It is then clear that the states on the unsplit sites
decompose into a single tensor product of irreducibles $R$ and
$\bar R$ as claimed. The string operators again correspond to the
difference in grading between that natural in $\cal F$ and our
choice, which agrees with the natural one in ${\cal F}'$. With our
choice, the (anti-)commutation relations obeyed by the generators
of the larger {\osp} SUSY are consistent with the stated grading,
as discussed in more detail in App.\ \ref{reposp}.

An advantage of the unconstrained representation of the states on
the split sites is that it makes the pure limit $p=0$ transparent.
The pure Ising problem in the anisotropic time-continuum limit in
this representation gives a nearest-neighbor ``hopping''-type
Hamiltonian for fermions and bosons, which is a sum over $i$ of
$\NSi$ and $\XSi$, with coefficients. This is a lattice version of
the Dirac fermion and its SUSY partner, the so-called
$\beta$-$\gamma$ system of bosonic ghosts. However, in the general
disordered case, there are additional terms which we expressed
previously using the zeroth fermion. In App.\ \ref{reposp}, we
show how the $G_i\bar{G}_i$ terms in the Hamiltonian can be
expressed in the present language. The other term, which becomes
$\NSi'$ on the NL, is much more difficult to express in this
language, and we return to this problem in Sec.\ \ref{general}.

A slight subtlety involved in the definition of the complex bosons
is the following. If we express them in terms of the $a$ bosons,
which diagonalize the form $\Nbi$, Eq.\ (\ref{diagonal_form}), we
obtain a singular Bogoliubov rotation: \beq b_{\mu i} =
\frac{\aph_{\mu i} + \abd_{\mu i}}{\sqrt 2}, \qquad \bb_{\mu i} =
\frac{\ad_{\mu i} - \abph_{\mu i}}{\sqrt 2}. \label{Bogrot} \eeq
The singularity of this transformation is seen in the fact that
the formal expression for the $b$, $\bar b$ vacuum (on a single
unsplit site $i$), defined by $b|{\tilde 0}\ra={\bar b}|{\tilde
0}\ra=0$, is \beq |{\tilde 0}\ra \propto \exp
\bigl(-\sum\nolimits_{\mu} \ad_{\mu i}\abd_{\mu i} \bigr)|0\ra,
\eeq where $|0\ra$ is the $a$, $\bar a$ vacuum defined in Eq.\
(\ref{defvac}), and leads to a series for the squared norm $\la
{\tilde 0}|{\tilde 0}\ra$ which is not convergent. A way out of
this problem is to regularize the Bogoliubov rotation
(\ref{Bogrot}) as follows: \bea b_{\mu i} &=& \cos\phi \,
\aph_{\mu i} + \sin\phi \, \abd_{\mu i}, \no\\ \bb_{\mu i} &=&
\sin\phi \, \ad_{\mu i} - \cos\phi \, \abph_{\mu i},
\label{Bogrotreg} \eea where $\phi = \pi/4 - \omega/2$ with $0 <
\omega \ll 1$. With such regularized transformation, the $b$,
$\bar b$ vacuum \beq |{\tilde 0}\ra = \frac{1}{\cos^n \phi} \exp
\bigl(-\tan \phi \sum\nolimits_{\mu} \ad_{\mu i}\abd_{\mu i}
\bigr)|0\ra \label{vacnorm} \eeq is well-defined and normalized to
1.

If we now use the regularized relations (\ref{Bogrotreg}), the
expression for $\Nbi$ becomes (to first order in $\omega$) \beq
\Nbi
=
\bd_{\mu i} \bbd_{\mu i} + \bbph_{\mu i} \bph_{\mu i} + \omega
(n_{bi} + n_{\bb i}) - n, \eeq with bosonic number operators
defined as \beq n_{bi} = \bd_{\mu i} \bph_{\mu i}, \qquad n_{\bb
i} = -\bbd_{\mu i} \bbph_{\mu i}. \label{bosnumber} \eeq The
fermionic sector in our formulation is finite dimensional, and
there are no similar problems with the fermions. However, to
maintain the exact cancellations between fermions and bosons, we
will modify the definition of $f$ and $\fb$ similarly to that for
the bosons. One effect of this is that the supersymmetric analog
of Eq.\ (\ref{vacnorm}) lacks the factor $1/\cos^n\phi$. Then the
{\os}-invariant combination (\ref{SUSYforms}) takes the form \bea
\NSi &=& \fd_{\mu i} \fbd_{\mu i} + \fbph_{\mu i} \fph_{\mu i} +
\bd_{\mu i} \bbd_{\mu i} + \bbph_{\mu i} \bph_{\mu i} \br +
\omega(n_{fi} + n_{\fb i} + n_{bi} + n_{\bb i}), \label{NSireg}
\eea where the fermionic number operators are defined in a natural
way: \beq n_{fi} = \fd_{\mu i} \fph_{\mu i}, \qquad n_{\fb i} =
\fbd_{\mu i} \fbph_{\mu i}. \eeq The term first-order in $\omega$
breaks the SUSY down to gl($n|n$), which is still enough SUSY to
ensure cancellation of fermions and bosons.

We can make this term appear more natural by the following
considerations. It is a regularizer which suppresses contributions
to the partition function from high fermion and especially boson
numbers on any site. We can introduce it in a more symmetric way
by inserting $\exp\sum_i\omega(n_{fi}+n_{\fb i} +n_{bi}+n_{\bb
i})$ between all the $T_{1i}$'s and $T_{2i}$'s in the partition
function; to first order in $\omega$, the effect is the same. Such
an insertion is a precaution similar to that often used in network
models and nonlinear sigma models of localization. The $\omega$
term represents a non-zero imaginary part of the frequency in
those problems, and as in the present case breaks the symmetry to
a subgroup. In the superspin chain language, the operator which
$\omega$ multiplies is one component of the staggered
magnetization, the order parameter for the chain. The term, with
$\omega\to 0$, is used just in case this develops a spontaneous
expectation value, since it picks a direction for the ordering in
superspin space and cuts off infrared divergences. Note that the
state with each site in the vacuum state for the $f$'s, $\fb$'s,
$b$'s, $\bb$'s is the N\'{e}el state corresponding to such order,
and is invariant under the subalgebra gl($n|n$). The
symmetry-breaking term will be important in Sec.\ \ref{lowtemp}.

\section{Dimerized limit and the one-dimensional case}
\label{lowtemp}

This Section lies somewhat outside of the main line of our
development; the latter continues in Sec.\ \ref{general}. Here we
consider our model in the vicinity of the NL deep in the
low-temperature phase. In terms of the superspin chain with the
Hamiltonian (\ref{ham2}), in this phase we have $h \ll k$. Then in
the zeroth approximation we may neglect the $h$ couplings
completely. Then the chain (\ref{ham2}) is broken into
disconnected pairs of superspins. The Hamiltonian for one such
pair is \beq H_k =  4 \str \La {\bar G} \La G = 4 \str \La G \La
{\bar G}, \label{ham4} \eeq where the coupling constant $k$
(overall energy scale) was taken to be equal to 4 for later
convenience, and we used the cyclic property of the supertrace. We
can try to solve this Hamiltonian and hope to infer some
information about the low temperature phase of our original model.
However, we wish to sound a note of caution: we are considering a
certain double limit of the original lattice Ising model, first
the anisotropic limit, then the ``low $T$'' limit, $h\ll k$. It is
not entirely clear that this really represents the low $T$ limit
of the nearly isotropic Ising model, where we pass to low $T$
close to the NL, and perhaps then go to the anisotropic limit.

We will make use of the realization of the representations $R$ and
$\bar R$ in Fock spaces of unconstrained fermions and bosons. For
simplicity we will work out the details for osp($3|2$) only. In
this case the necessary construction of $R$ and $\bar R$ and the
invariant products of superspins is given in Appendix
\ref{reposp}. {}From it we obtain \beq H_k = \lam J - J^2,
\label{Hamk} \eeq with \beq J = \fd\fbd + \fb f + \bd\bbd + \bb b.
\label{J} \eeq

We anticipate that the eigenstates of the Hamiltonian $H_k$ may
have arbitrarily large bosonic occupation numbers, and we may
encounter convergence problems typical in such cases. These are
avoided, however, if we remember the $\omega$ term, discussed in
Sec.\ \ref{unconstr}. As explained there, it plays the role of a
symmetry-breaking regulator that picks a direction for ordering,
similar to $S_1^z - S_2^z$ for the problem of two
antiferromagnetically coupled su(2) spins. Thus, we add to our
Hamiltonian the term \beq H_\om = \om(\nf + \nb + \nfb + \nbb).
\eeq

The resulting Hamiltonian \beq H = H_k + H_\om \label{ham5} \eeq
is {\it identical\/} to the one studied by Balents and Fisher in a
one-dimensional localization problem (see Eq.\ (3.31) in Ref.
\onlinecite{bf}). This is the problem of spinless fermions on a 1D
lattice with random hopping amplitudes described by the
Hamiltonian \beq {\cal H} = - \sum_n t_n \bigl(\cd_n \cph_{n+1} +
\cd_{n+1} \cph_n \bigr). \eeq The continuum limit of this model
gives left and right moving spinless Dirac fermions with random
mixing between them: \beq {\cal H}_c = \int \!\! dx \Psi^{\dagger}
\bigl(- i \sz \dx + V(x) \sy \bigr) \Psi, \label{contham} \eeq
where $\Psi(x)$ is a two-component spinor field, $\sigma^i$ are
Pauli matrices, and the random potential is Gaussian with non-zero
mean and variance: \bea \disave{V(x)} &=& V_0,  \no\\
\disave{\bigl( V(x)-V_0 \bigr) \bigl( V(x')-V_0 \bigr)} &=&
2D\de(x-x'). \eea

The generating functional for the Green's functions of this
Hamiltonian at a given energy $\eps + i\eta$ may be
supersymmetrized in the standard way. After disorder averaging the
$x$ coordinate may be interpreted as imaginary time, and the two
components of the fermion can be viewed as labeling two sites,
which correspond to our split sites. This leads to an effective
quantum Hamiltonian, which is exactly given by Eq.\ (\ref{ham5})
with \beq \lam = V_0/D, \qquad \om = \eta - i\eps. \eeq

The 1D model with the Hamiltonian (\ref{contham}), and related
models, have a long history, and most of the relevant work is
concisely summarized in Ref.\ \onlinecite{mckenzie}. In
particular, the density of states for this problem was found for
$\lam = 0$ in 1953 by Dyson \cite{dyson}, and for arbitrary $\lam$
by many authors \cite{1d}. The mathematically equivalent problem
of diffusion in a 1D random medium was studied by Bouchaud {\it et
al\/} \cite{bcgl}. The density of states $\rho(\eps)$ behaves at
small energies as \bea \rho(\eps) &\propto& \frac{1}{\eps |\ln^3
\eps|}, \quad  \lam = 0, \\ \rho(\eps) &\propto& \eps^{\lam - 1},
\quad  \lam > 0. \label{dos} \eea

In the superspin language the density of states $\rho(\eps)$ is
related to the expectation value of some operator in the ground
state of the Hamiltonian (\ref{ham5}) \cite{bf}. Namely, it is
proportional to the staggered component $h_2 - {\bar h}_2$ of the
superspin (see Appendix \ref{reposp}) \beq \rho(\om) \propto \la 1
- \nf - \nfb \ra \propto \om^{\lam - 1}. \eeq This quantity
measures the amount of the symmetry breaking in the ground state
of two superspins. {}From the last equation it follows that the
symmetry is spontaneously broken on the NL (which is a point in
the 1D model) and below it. Moreover, below the NL, where $\lam <
1$, the density of states (the order parameter of the spin chain)
diverges as $\om \to 0$. On the NL it is constant, and above the
NL it vanishes as a power of $\om$.

Because the SUSY representations are the same, we have in fact
shown that in the 1D off-diagonal disorder problem, {\em there is
a larger SUSY} {\osp} {\em at the point $\lambda=1$}. This has not been
noticed previously to our knowledge. This suggests that such
Nishimori points, lines, etc, may be common in some classes of
random fermion problems. We also note here that in the 1D
classical RBIM, which of course has no finite $T$ phase
transition, there is a Nishimori point at which the correlation
identities Eqs.\ (\ref{correq}) hold. That problem can be
represented using fermions on one unsplit site with the $T_{1i}$
transfer matrices only, which are of the $h$-coupling type, in
contrast to the model considered here, and is easily solved in
this language.

\section{Final representation and the generalized model}
\label{general}

In this Section we continue the general consideration of the RBIM
problem. Here we focus our attention on the NL, that is, we
consider the \osp-invariant Hamiltonian (\ref{ham3}). First we
analyze and solve the problem of finding a way to describe the
term $\Nfi$ in the spaces $R$, $\bar R$ on the split sites. The
problem is solved by using another representation in a space
${\cal F}''$, and the spaces can be viewed as representations of a
larger SUSY algebra, \Osp. Using only terms of the form of the two
couplings we have already seen, we then introduce a more general
nearest-neighbor superspin chain, and discuss its phase diagram,
for reference in the following Sections.

First let us note that Eq.\ (\ref{ham3}) is somewhat schematic.
Let us again consider the fermionic replica formalism with $n \to
0$ instead of SUSY. The term ${\NSi}'$ is then replaced by
${\Nfi}'$. According to our general discussion of how to map
operators in ${\cal F}$ into ${\cal F}'$ (see Sec.\
\ref{enhsusy}),  \beq \Nfi' = i\eta_{ai}\xi_{ai} + n + \frac{1}{2}
\label{nfi} \eeq is correct as it stands in ${\cal F}'$. Even
though the operator ${\Nfi}'$ is perfectly legitimate, it does not
admit any simple expression in terms of the so($2n+1$) generators
$G_i$ and ${\bar G}_i$. We would like to write it as a sum of
products of operators in the spinor representations $R$, $\bar R$
on the split sites. Of course, individual fermion operators
$\xi_{ai}$, $\eta_{ai}$ do not commute with the constraint, and
cannot be used. Instead they must be replaced by ${\Bbb
Z}_2$-invariant operators. As explained in Sec.\ \ref{enhsusy}, we
can find operators in ${\cal F}'$ with the anticommutation
properties of the fermions in $\cal F$ for the components {\em
other than the zeroth}. These are fermion bilinears times a
string; see Eq.\ (\ref{subst}). A general proof that it is
impossible to find a set of operators with the anticommutation
relations of the {\em full} set of real fermion operators
$\eta_{ai}$ and $\xi_{ai}$ in the space ${\cal F}'$ is to notice
that they should form a Clifford algebra with $2N(2n+1)$
generators, where $N$ is the number of unsplit sites in the chain.
This Clifford algebra has a single non-trivial representation of
dimension $2^{N(2n+1)}$. This space is the same as an
unconstrained Fock space for $2n+1$ complex fermion operators at
each unsplit site, i.e. ${\cal F}'$ but without the constraints.
The total number of states in ${\cal F}'$ is only $2^{2Nn}$,
because of the $N$ constraints. So the operators we require cannot
have the anticommutation relations of free fermions for all the
sites. Indeed, in our grading on ${\cal F}$, single fermion
operators are even, and so would be expected to obey commutation
relations from a SUSY point of view. In ${\cal F}'$, there are
corresponding fermion bilinears, like Eq.\ (\ref{subst}) but
without strings, and these do commute on different sites (as
mentioned already in Sec.\ \ref{enhsusy}).

We can also try the unconstrained representation. Then again, we
can represent each of the $2n$ real fermions $\eta_{\al i}$ and
$\xi_{\al i}$ on each split site using Eqs.\ (\ref{ffermions}),
and the resulting Clifford algebra for $2N$ split sites yields the
correct number of states. This description carries over easily to
the SUSY version. But the above proof shows that no matter what
strings or other factors we introduce into a construction of
operators, we cannot produce the anticommutation relations for
$2n+1$ real fermions at each split site, and we are no nearer
writing $\NSi'$ as a product of simple expression in $R$ and $\bar
R$. What we would have to do is map the problematic part $n_{0i}$
of the operator back from ${\cal F}'$ to ${\cal F}$. Because of
the constraint in the former space, the resulting operator (still
in the fermionic replica formalism) must equal 1 when $\Nfi$ is
odd, 0 when $\Nfi$ is even, or similarly for $\NSi$ in the SUSY
formalism. It is not clear how we would write this as a coupling
of the two split sites at $i$.

There is nonetheless a way out of this problem, motivated by the
following observation. If we consider a single spinor
representation $R$ of so($2n+1$) (thus, in the fermionic replica
formalism once more), then it is in fact possible to find
operators with the anticommutation relations of the real fermions
$\xi_a$. These are the generators of a Clifford algebra with an
odd number $2n+1$ of generators, which has an irreducible
representation of dimension $2^n$ (the familiar $2\times 2$ Pauli
matrices are the case $n=1$). The commutators $ \frac{i}{2}[\xi_a,
\xi_b]$ of these operators are the generators of so($2n+1$), as we
have already seen. The operators $\xi_a$ transform as a vector of
so($2n+1$). If we now consider these so($2n+1$) generators
together with the $\xi_a$ (divided by $\sqrt{2}$), then we may use
the fact that the {\em commutators} of these operators (or
matrices) together obey the relations of the generators of
so($2n+2$), and the spinor $R$ can be identified with one of the
two distinct irreducible spinor representations of dimension $2^n$
of so($2n+2$). This construction can also be applied to the
representations $\bar R$ [which for so($2n+1$), though not for
{\osp}, is isomorphic to $R$].  This construction also extends
easily to the many-site problem, {\em by taking the operators now
replacing $\xi_a$ to commute on different sites}. We may therefore
write down our term $\Nfi'$ as a sum of products of bilinears of
these operators, and this can also be extended to the SUSY
construction, using operators with the relations of $q_\alpha$,
$r_\alpha$, on each site, but which anticommute on different
sites.

Thus, we have learned that our spaces of states $R$, $\bar R$ at
alternate sites can be viewed as irreducible spinor
representations of {\Osp} (but note that this algebra is not a
symmetry of our Hamiltonian or transfer matrices so far). There
are two inequivalent lowest-weight spinor representations $R_e$,
$R_o$ (for ``even'' and ``odd'') of \Osp, in which all states can
be assigned positive norm-squares. We can identify $R$ with, say,
$R_e$. Similarly, $\bar R$ can be identified with a highest-weight
spinor ${\bar R}_e$, which is dual to $R_e$, and in which the
inner product is indefinite, since as we have seen states with an
odd number of $\bb$ bosons have negative squared norms.

Viewing the spaces of states in this way, we can give yet another
explicit construction, with which we can finally write the
operators $\Nfi'$ and $\NSi'$ in a simple way. It is convenient to
keep much of the notation the same as before. We introduce
additional complex fermions $f_{0i}$, $\fb_{0i}$ to the set
$f_{\mu i}$ of Eq.\ (\ref{ffermions}), and define a space ${\cal
F}''=\otimes_i{\cal F}''_i$ consisting of the states on the split
sites with an even number of fermions plus bosons: \bea n_{bi} +
n_{fi} + \fd_{0i} \fph_{0i} &=& \text{even},
\\ n_{\bb i} + n_{\fb i} + \fbd_{0i} \fbph_{0i} &=& \text{even}.
\eea It is clear that such states are in one-one correspondence
with those in the unconstrained representation. The construction
of the correspondence is similar to that for the states in the
spaces ${\cal F}$ and ${\cal F}'$ in Sec.\ \ref{enhsusy}. All the
states can be obtained from the vacuum, which is the lowest-
(highest-) weight state in $R_e$ (${\bar R}_e$), by the action of
the bilinears in the creation operators. Then in addition to Eq.\
(\ref{ffermions}) we also define \beq f_{0i} = \frac{\xi_{0i} + i
\xi'_{0i}}{\sqrt 2}, \qquad \fb_{0i} = \frac{i \eta_{0i} +
\eta'_{0i}}{\sqrt 2}. \eeq Now the so($2n+2$) generators on a
single site are replaced by \beq \frac{i}{2}[\xi_{ai}, \xi_{bi}],
\qquad \frac{i}{2}[\xi'_{0i}, \xi_{ai}], \label{o2gens'} \eeq
where the first set, again, spans the subalgebra so($2n+1$), and
the second set transforms as a vector under this subalgebra. There
are similar expressions for ${\bar R}_e$. We emphasize that the
operators $\xi_{ai}$, $\xi_{0i}'$, $\eta_{ai}$, $\eta_{0i}'$ obey
canonical anticommutation relations, while $q_{\alpha i}$,
$r_{\alpha i}$ obey canonical commutation relations, of the same
form as in Eqs.\ (\ref{anticomm}), (\ref{commrel}), (with negative
norm states appearing in connection with the $r$'s, see Eqs.\
(\ref{bbosons}) and following). Our choice of grading is again
equivalent in the constrained subspaces ${\cal F}''_i$ to their
natural grading as subspaces of Fock spaces. Finally, in the
representation in ${\cal F}''$, the operator $\NSi'$ undergoes the
replacement \bea \NSi' &=& i \eta_{ai} \xi_{ai} - r_{\al i} \jab
q_{\be i} + \frac{1}{2} \no\\ &\mapsto& 2
\eta'_{0i}\eta_{ai}\xi_{ai}\xi'_{0i} + 2 i \eta'_{0i} r_{\al i}
\jab q_{\be i} \xi'_{0i} + \frac{1}{2}. \eea These results may
also be established by passing directly from the (averaged)
unconstrained representation to the final representation, by using
a substitution similar to Eq.\ (\ref{subst}), but applied here to
the split sites.

We can now organize the generators of {\Osp} in superspins,
similar to Eqs.\ (\ref{suspin}, \ref{suspinb}), and including the
additional odd generators:
\bea G' \! &=& \! \left( \!
\begin{array}{ccc}
        0 & \xi'_0 \xi_b & \xi'_0 q_\be \\
        \xi_a \xi'_0 &
        \xi_a \xi_b - {1 \over 2} \de_{ab} &
        \xi_a q_\be \\
        i J_{\al \ga} q_\ga \xi'_0 &
        i J_{\al \ga} q_\ga \xi_b &
        i J_{\al \ga} q_\ga q_\be - {1 \over 2} \dab
        \end{array} \! \right) \! ,
\label{superspin} \\
{\bar G}' \! &=& \! \left( \! \begin{array}{ccc}
        0 & \eta'_0 \eta_b & \eta'_0 r_\be \\
        \eta_a \eta'_0 &
        \eta_a \eta_b - {1 \over 2} \de_{ab} &
        \eta_a r_\be \\
        i J_{\al \ga} r_\ga \eta'_0 &
        i J_{\al \ga} r_\ga \eta_b &
        i J_{\al \ga} r_\ga r_\be - {1 \over 2} \dab
        \end{array} \! \right) \! .
\label{superspinb} \eea Note that these {\Osp} superspins contain
the original {\osp} superspins $G$ and $\bar G$ as submatrices.
The odd generators are those containing an odd number of fermion
operator factors, or equivalently an odd number of boson operator
factors. With the help of the {\Osp} superspins, both terms in the
Hamiltonian (\ref{ham3}) may be written in a unified way: \bea h
\NSi' &=& \lim_{k\to 0} \left( \str C G'_i C {\bar G}'_i +
\frac{h}{2} \right), \\  k \str {\bar G}_i G_{i+1} &=& \lim_{h\to
0} \left( \str C {\bar G}'_i C G'_{i+1} + \frac{h}{2} \right),
\label{kcoupl} \eea where we have introduced a $4n+2$-dimensional
diagonal matrices of coupling constants \beq C \equiv C(h,k) =
\text{diag}(h k^{-1/2}, k^{1/2} \openone_{4n+1}), \eeq and the
supertrace $\str$ in this space is defined in the same way as the
previous $\str$. These two terms represent two different
\osp-invariant products of two subsets of the {\Osp}generators. It
should be clear that the representation in ${\cal F}''$ can also
be used off the NL, by giving certain terms different
coefficients.

It is now natural to consider a generalized Hamiltonian
\bea
H = && \sum_i \bigl(\str C(h_A, k_A) G'_i C(h_A, k_A) {\bar G}'_i  \br
 \quad + \str C(h_B, k_B) {\bar G}'_i C(h_B, k_B) G'_{i+1} \bigr) \no \\
&+& \frac{h_A + h_B}{2} N, \label{hamgen} \eea parametrized by
four coupling constants. In such a Hamiltonian both types of
\osp-invariant couplings appear on every bond between the split
sites. In addition, they are staggered between the two sublattices
of bonds $A$ and $B$ of our chain. Our NL Hamiltonian (\ref{ham3})
is a particular extreme limit, where on alternate bonds one or the
other coupling is zero. It is obtained from Eq.\ (\ref{hamgen})
for the special values of the parameters \beq h_A = h, \quad k_A =
0, \qquad h_B = 0, \quad k_B = k. \eeq We believe it may be
helpful to consider these more general models, since they are so
closely related to that for the RBIM, and use only couplings that
appear anyway in the RBIM case; however, we emphasize that {\em it
may not be possible to obtain these models as anisotropic limits
of random fermion or network models}. When \beq h_A = k_A, \qquad
h_B = k_B, \label{Ospcond} \eeq the model is invariant under the
whole of {\Osp}. We should note that in principle we can also
consider this generalization in the discrete imaginary time model,
and also off the NL, where however the breaking of the symmetries
would lead to twice as many parameters. The additional parameters
would generalize $\lambda$, $\lambda^*$ in Sec.\ \ref{supspin},
and there would be one for each of $k_A$, $k_B$, $h_A$, $h_B$, a
total of eight parameters in the Hamiltonian. In particular,
another model due to Cho and Fisher\cite{cf} fits into this
general description, as we will see in Sec.\ \ref{CFm}.

\begin{figure}
\epsfxsize=3.4in \centerline{\epsffile{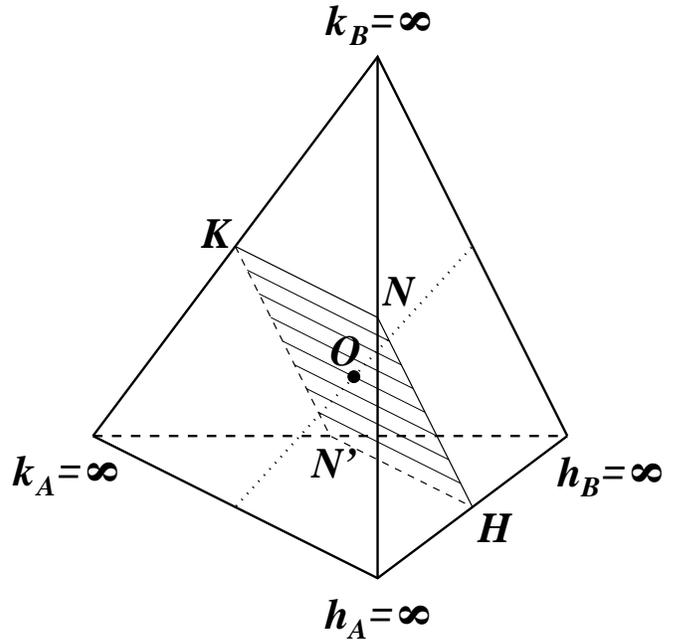}}
\vspace{0.25in} \caption{A possible phase diagram of the
generalized Hamiltonian (\protect\ref{hamgen}), discussed in
detail in the text.} \label{fig3}
\end{figure}

The Hamiltonian (\ref{hamgen}) contains four parameters, but since
the overall energy scale is unimportant, the phase diagram can be
plotted in terms of the three independent ratios of parameters. We
will consider only positive values of all couplings, though
negative values may also give well-defined models. The phase
diagram can then be drawn in a symmetrical manner in a
three-dimensional tetrahedron, as a portion of projective space
(see Fig.\ \ref{fig3}). Each face of the tetrahedron is defined by
one of the four parameters vanishing. The opposite vertex is where
that parameter goes to infinity, or equivalently the other three
all go to zero.

The edges of the tetrahedron correspond to models with two
vanishing couplings. For example, the vertical edge connecting the
vertices $h_A = \infty$ and $k_B = \infty$ represents the
Hamiltonian (\ref{ham3}) for the RBIM on the NL (to avoid
confusion, recall that the whole discussion is a generalization of
the NL, since all the models have the larger {\osp} SUSY). There
is another such line represented by the horizontal edge connecting
the vertices $h_B = \infty$ and $k_A = \infty$. These two
Hamiltonians are related by a reflection through a lattice site.
Such an operation is thus a symmetry of the whole diagram, which
interchanges $A$ with $B$. The line $k_A=k_B$, $h_A=h_B$ is
invariant under this operation, and the operation acts as a
180$^\circ$ rotation about this line. On each NL, there is a
multicritical point, $N$ and its image $N'$.

The edges where both non-zero couplings are on the same sublattice
of bonds (e.g.\ $A$) represent the two extreme cases of
fully-dimerized chains, which have a gap in their energy spectrum.
By analogy with other antiferromagnetic (super-)spin chain models,
we expect that the regions adjacent to these lines are also gapped
phases. There must be at least one phase transition between these
two extremes. One way to see this is to consider a chain with open
ends, and an even number of split sites. In one phase the dimers
extend all the way to the ends of the chain, in the other a single
superspin is left unbonded with a neighor at each end. This
corresponds to a chiral edge degree of freedom in the 2D lattice
model. A phase transition must occur to change the number of such
boundary spins or edge channels, assuming these survive off the
edges of the tetrahedron. We will assume that, as expected on the
NL in the RBIM, there is a single transition between the two
phases. Then there must be a phase-boundary surface between those
two edges, indicated schematically (since its exact position is
unknown) by the shaded surface in Fig.\ \ref{fig3}. The points $N$
and $N'$ are two vertices of this rhomboidal surface, which also
contains the line of reflection symmetry. However, we note that an
intermediate phase, in place of some portion of the critical
surface, is also possible, though this is not expected on the NL
in the RBIM.

The two other edges of the tetrahedron are where either only the
$k$'s are nonzero, or only the $h$'s, and we denote these models
``$k$-only'' and ``$h$-only''. They intersect the phase boundary
(if there is a unique transition on these edges) at points labeled
$K$ and $H$ in Fig.\ \ref{fig3} (no confusion should result from
this notation). In these models, the reflection symmetry and the
assumption of a single transition implies that $K$ is $k_A=k_B$,
and $H$ is $h_A=h_B$ (and other parameters zero).

The tetrahedral phase diagram also contains the line given by Eq.\
(\ref{Ospcond}) where the generalized Hamiltonian (\ref{hamgen})
has the {\Osp} symmetry. This line, shown dotted in Fig.\
\ref{fig3}, intersects the critical surface at a critical point
$O$ (black dot), where all four couplings are equal. Again, this
point is unique if we assume there is a single transition on this
line; it is $k_A=k_B=h_A=h_B$.

\section{Cho-Fisher, $\lowercase{k}$-only, and $\lowercase{h}$-only models}
\label{CFm}

In this Section we consider a model studied numerically by Cho and
Fisher in Ref. \onlinecite{cf}. This is a network model, similar
to the Chalker-Coddington network \cite{cc} describing the integer
quantum Hall transition, but with only real matrices, and was
intended to represent the RBIM problem. We show that the Ising
model can be represented exactly as a network model, and that the
Cho-Fisher model does {\em not} represent the RBIM. Instead, it
can be mapped to some of the generalized Hamiltonians {\em
without} enhanced SUSY, introduced in Sec.\ \ref{general}.

The Cho-Fisher model is a network model, intended to capture the
universal aspects of the point $N$ in the RBIM, which can be
viewed as a generalization of the 1D model discussed in Sec.\
\ref{lowtemp}. It was constructed as a generalization of the model
whose action is given in Eq.\ (\ref{contham}), in which the two
components of the fermion are replaced by any number of sites in a
1D chain, with random nearest-neighbor hopping that generalizes
the $\sigma^y$ term in Eq.\ (\ref{contham}), and, in general,
different parameter values $V_0$ and $D$ on alternate bonds in the
chain. Then use of replicas or SUSY to perform the disorder
average leads to a generalization of the quantum (super-)spin
Hamiltonian $H_k+H_\omega$, Eq.\ (\ref{Hamk}) in Sec.\
\ref{lowtemp} (we will disregard here the regularising term
$H_\omega$), in which the same form of coupling appears for each
pair of nearest neighbors, but with the coefficient of $H_k$
taking two values $k$, $k'$ on alternate bonds, and similarly for
$\lambda$, $\lambda'$. We emphasize that at this stage we are
using the unconstrained representation of the space of states of
the chain. Cho and Fisher specialized to the case $ k=k'$ (i.e.\
node independent disorder strength), and went back from their
time-continuum model to a discrete-time (network) model, similar
to that in Ref.\ \onlinecite{cc}, in order to perform numerical
calculations. They claimed that their network model has a
multicritical point in its phase diagram with critical properties
remarkably similar to those of the multicritical point on the NL.
In particular, the critical exponents along the two scaling axes
near the multicritical point were found to be fairly close
numerically to the ones known for the point $N$ in the RBIM from
the high-temperature expansion of Singh and Adler \cite{sa}. Also,
simulation in Cho's thesis \cite{cf} of a network model that
corresponds precisely to the RBIM (as we will explain) gave
similar values.

\begin{figure}
\epsfxsize=4in \vspace{-1.4in}
\centerline{\epsffile{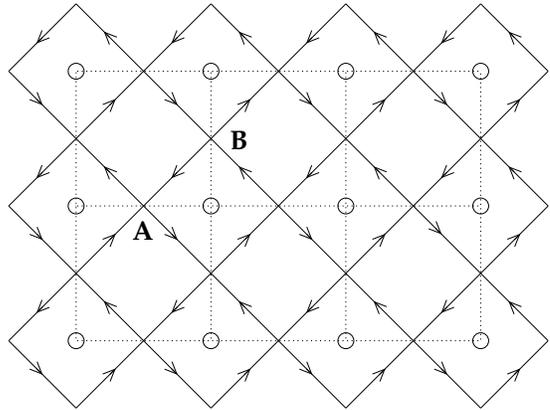}} \vspace{-1.3in}
\caption{Relation of the Ising model and the network model. Ising
spins are located at the open circles, and bonds are shown dotted.
Solid lines with arrows form the ``medial graph'', on which the
network model is defined. Examples of nodes on each of the two
sublattices, corresponding to the horizontal and vertical bonds,
are labeled $A$, $B$, respectively. In this paper, we consider
periodic boundary conditions in the vertical direction, and free
in the horizontal, as shown here.} \label{network}\vspace{-0.1in}
\end{figure}

Now, we wish to point out that it is possible to relate network
models more directly to the transfer-matrix formulation for
fermions, as in the Ising model. First we describe the network
models. A portion of the network is shown in Fig.\ \ref{network},
where the solid lines with arrows are where the particles
propagate. The particles propagate in discrete time, at each time
step moving to the next link in the ``forward'' direction shown by
the arrows, and therefore turning either left or right at each
time. The evolution is described by a unitary S-matrix which gives
the amplitudes for turning either right or left at each node
\cite{cc}. This can be replaced by a one-particle transfer matrix,
which adds one row of nodes to the system, evolving the
wavefunction of the particle upwards in the figure. In the
Cho-Fisher model, the one-particle transfer matrix for one node
has the form \beq M=\left(\begin{array}{cc}
 \cosh \theta&\sinh\theta\\ \sinh \theta & \cosh
 \theta\end{array}\right).\eeq
The parameter $\theta$ for each node is random, taking values
$\pm|\theta|$ with probabilities $1-p$, $p$, independently. Also,
the magnitudes of $\theta$ can be staggered, taking different
values $|\theta_A|$, $|\theta_B|$ on the two sublattices of nodes
labeled $A$, $B$ in Fig.\ \ref{network}. The so-called isotropic
case is where $\sinh|\theta_A|\,\sinh|\theta_B|=1$. This leaves a
one-parameter family of models; in the original network model
\cite{cc}, a transition occurred when $|\theta_A|=|\theta_B|$, the
``self-dual'' point.

To exhibit a relation with the Ising model transfer matrices, we
use a second-quantized formulation of the network, as a
noninteracting fermion field theory. The evolution in the
imaginary-time (vertical) direction is described by a transfer
matrix constructed from the one-particle one. We can write this by
drawing on earlier work \cite{grs}. Though the latter was on a
different model, the basic Eq.\ (2) in that work is applicable for
{\em any} transfer matrix, with matrix elements \beq
\left(\begin{array}{cc}
 \alpha&\beta\\ \gamma& \delta\end{array}\right).\eeq
Thus in our case, $\alpha=\delta=\cosh\theta$,
$\beta=\gamma=\sinh\theta$. Using only one species of fermions,
and dropping the bosons in Eq.\ (2) of Ref.\ \onlinecite{grs}
since we will not be averaging here, we replace $f_1$ in Ref.\
\onlinecite{grs} by $f$, $f_2$ by $\fb'$, (where $f$ and $\fb'$
obey canonical anticommutation relations) and obtain \beq
V=:\exp[\tanh\theta(f^\dagger\fb'+\fb'^\dagger f)]:
(\cosh\theta)^{n_f+n_{\fb'}}.\eeq Here the colons $:\ldots:$
indicate normal ordering with destruction operators to the right.
Then, after making the particle-hole transformation
$\fb^\dagger=\fb'$, we can prove the identity \beq
V=e^{\theta(f^\dagger\fb^\dagger + \fb f)},\eeq by verifying that
all matrix elements of the two expressions are equal. But this now
has the form of the fermionic representation of the squared Ising
model (i.e., $n=1$), as in Eqs.\ (\ref{tvth}), (\ref{SUSYforms})
(dropping the bosons in the latter), up to constant factors in the
vertical case. This means that the split sites on a single row
correspond to one row of links of the network. The relation of the
original Ising lattice and the network model is as shown in Fig.\
\ref{network}; in particular, the two sublattices of nodes $A$,
$B$, correspond to horizontal and vertical bonds respectively. The
relationship of second-quantized transfer matrices holds true for
arbitrary values of $\theta$ at each node, and also remains true
when bosonic partners are introduced in preparation for averaging.

The important corollary to this is that for the transfer matrices
of the sort appropriate for the {\em horizontal} bonds (labeled
$A$ in Fig.\ \ref{network}), we have \cite{chothes} $2K=\theta$.
The Cho-Fisher network model takes the parameter $\theta$ at the
nodes to have independent random signs. Hence it is {\em
precisely} equivalent to the use of transfer matrices (\ref{Tv}),
(\ref{Th}), with the binary distribution of the type (\ref{PK})
for {\em both} the horizontal couplings $K_{\bi,\bi + \bx}$ and
the {\it dual} ${\tilde K}_{\bi,\bi + \by}$ to the vertical
couplings. Note that the Cho-Fisher model is not in fact
isotropic, even when the (magnitudes of the real parts of the)
$K$'s, and the probabilities $p$, on the horizontal and vertical
links are the same, which is what we termed isotropic above; this
is because of the way the random signs are introduced. Another
popular parametrization for the network models uses the S-matrix
at each node \cite{cc}, where the S-matrix is a real orthogonal
matrix in the present case, with one of its off-diagonal matrix
elements (say, the amplitude for turning right) denoted
$t=\sin\phi$. In this case the equivalence to the Ising model
squared is $\tanh 2K=\sin \phi$ for the horizontal couplings.

Since negative dual couplings $\tilde K$ correspond to complex
Ising couplings $K$, the Cho-Fisher model does {\em not}
faithfully reproduce the RBIM with $\pm K$ couplings. Instead, in
replicated fermion language, both types of bond are represented
after averaging by transfer matrices of the horizontal type. One
might imagine that this is the $k$-only model, with parameters
$\lambda_A$, $\lambda_B$ included, so that the {\osp} SUSY is
present when $\lambda=\lambda'=1$. But in fact, carefully
following our mapping leads to different forms for the two bonds,
$A$, $B$. It is necessary once again to use the final (${\cal
F}''$) representation, and to pass to it directly from the
unconstrained representation. We find that, while the bonds
corresponding to the horizontal bonds in the Ising model involve
the real fermions $\xi_{0i}$, $\eta_{0i}$, in the $k$-coupling
terms as in Eq.\ (\ref{kcoupl}), for the vertical bonds those
fermion operators are replaced by $\xi_{0i}'$, $\eta_{0i}'$. For
$\lambda$ or $\lambda'=1$, these terms are invariant under an
{\osp} SUSY, but these are distinct {\osp} sub-superalgebras of
{\Osp} in the two cases, and so the Hamiltonian does not possess a
global {\osp} SUSY (though there is of course still {\os}). These
models therefore lie elsewhere in our space of fully-generalized
Hamiltonians with (in general) only {\os} SUSY.

We note here that by rotating the Cho-Fisher network by
$90^\circ$, we obtain after averaging (using the ${\cal F}''$
representation) a model which resembles the $h$-only model, but
again has different {\osp} SUSYs for the two types of bond.
Because taking the anisotropic limit usually does not affect the
universality class of the critical phenomena, the resulting spin
chain model should have the same critical phenomena as the one
described above for the Cho-Fisher model.

\section{Fixed points and non-linear sigma models}
\label{nlsm}

In this Section, we first speculate that a single fixed point, or
universality class controls much of the phase boundary in the
tetrahedral phase diagram. Then we discuss the nonlinear sigma
models that are related to our spin chains, and calculate, at weak
coupling, the RG beta functions for the coupling constants. The
results support the hypothesis of a flow towards the higher SUSY
as at point $O$ in the phase diagram. Finally, we discuss the
nonlinear sigma (and related) models for the more general, lower
SUSY ({\os}), or class D, random fermion problems.

\begin{figure}
\epsfxsize=3.4in \centerline{\epsffile{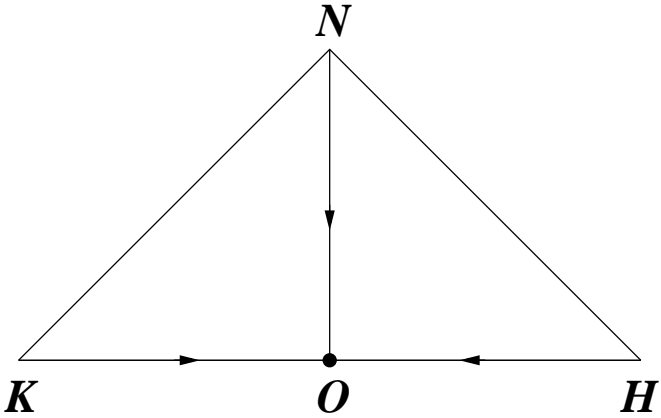}}
\vspace{0.25in} \caption{The critical surface of the generalized
Hamiltonian, after reduction using symmetry. The arrows indicate
our suggested RG flows to a fixed-point theory with the same
larger SUSY as the model at $O$.} \label{rgflow}
\end{figure}

For the generalized Hamiltonian with {\osp} SUSY, we argued (on
the assumption that there is a single transition surface) that the
phase boundary is a rhombus, and further there is a reflection
symmetry in the superspin chain, which, on this surface, acts as a
reflection. Then the phase boundary is a triangle, with points
$K$, $N$, $H$ at the vertices, as shown in Fig.\ \ref{rgflow}. The
point $O$, at which the model has the larger SUSY algebra {\Osp},
is now at the middle of one side of the triangle.

Under the RG, the {\Osp}-invariant model must flow to a critical
quantum field theory (presumably, a conformal field theory) which
also has the larger SUSY. Other models, represented by other
points in Fig.\ \ref{rgflow}, such as $N$, $K$, $H$ may flow to
some other fixed point theories of lower ({\osp}) SUSY, and it is
of course the fate of $N$ that concerns us in the RBIM model
problem. If we must make a guess as to the structure of the flows
and fixed points, the simplest guess is the one that involves the
fewest fixed points. Since there must be a fixed point theory with
{\Osp} SUSY, the simplest guess is then that the whole critical
surface flows to this fixed point. This is schematically
illustrated in Fig.\ \ref{rgflow} by the arrows, which are
intended to indicate that all models flow to the fixed point
corresponding to $O$ (note that the models at $N$, $K$, $H$, and
$O$ are {\em not} themselves fixed points of the RG). If correct,
this would imply that the critical exponents are the same at all
points in the critical surface shown in Figs. \ref{fig3} or
\ref{rgflow}. In particular, $N$, $K$, and $H$ would have the same
exponents. We can also imagine other scenarios in which $N$, $K$
and $H$ flow to a common fixed point, or to different ones, that
do not have {\Osp} SUSY. It is certainly possible that (one or
both) perturbations away from $O$ on the surface shown are
relevant; we are suggesting that they are both irrelevant. In the
absence of any understanding of the conformal field theory of the
{\Osp} or other fixed points in this system, we cannot prove or
disprove our suggestion. There are, however, other systems in
which an analogous effect occurs, as we will discuss below.

Now we introduce the nonlinear sigma models that should correspond
to the superspin chains, and should have transitions in the same
universality class or classes. First we utilize a standard
relation between antiferromagnetic spin chains and nonlinear sigma
models (see Refs.\ \onlinecite{affleck85} for a fairly general
discussion). We define an antiferromagnetic (super-)spin chain as
having an irreducible representation at each site, alternating
between some (say) lowest-weight representation $R$ and its dual,
say $\bar R$. The Hamiltonian should be something close to the
Heisenberg form which is the invariant bilinear form in the
generators of the symmetry algebra, with the antiferromagnetic
coupling that for a single pair of spins leads to the singlet
ground state (possible because we chose the dual representations).
Then the correspondence states that there is a nonlinear sigma
model with a certain target (super-)manifold, which can be
obtained from the representation $R$. The manifold is the same
coset space that appears as the coadjoint orbit of $R$, or in
coherent-state path integral constructions of $R$. Put simply,
this is the manifold swept out by acting on either a lowest- or
highest-weight of $R$ with all possible (super-)group elements.
The long-wavelength action of the nonlinear sigma model in $1+1$
dimensions contains only terms allowed by symmetry with two
derivatives. These comprise the usual ``kinetic'' type terms, and
also possible ``$\theta$-terms'' (this is not the same parameter
$\theta$ we used in Sec.\ \ref{CFm}). The derivation is controlled
by considering a sequence of representations $R$ with the lowest
weight going to infinity in the weight space, like the size of the
spin in SU($2$) going to infinity. Then the reciprocals of the
kinetic couplings have magnitude proportional to the lowest
weight, so the nonlinear sigma model is weakly coupled and
meaningful in the semiclassical limit. Also, in the absence of
staggering of the couplings in the spin chain, $\theta$ is
proportional to the lowest weight with a coefficient $\pi$ in a
suitable normalization (which is such that the bulk physics is
periodic when $\theta\to\theta+2\pi$). A $\theta$ term exists and
is nontrivial whenever the second homotopy group $\pi_2$ of the
target manifold is nontrivial. More generally, a $\theta$-term
involves a two-form on the target manifold (i.e.\ a magnetic field
for a charged particle moving on the manifold) that is invariant
under the symmetry, and this always exists in this construction
because it is part of the coherent-state construction of the
representation $R$ also. Noncompact factors in the manifold are
topologically trivial, but the term described always produces
boundary effects related to a boundary spin or edge states
\cite{xrs}. The nonlinear sigma model that results from this
correspondence in many cases has a phase transition at
$\theta=\pi$, when the target manifold has nontrivial $\pi_2$.

In our case, the target manifold for our general models on the NL
would be, for fermionic replicas, SO($2n+1$)/U($n$), or in the
SUSY formalism, OSp($2n+1|2n$)/U($n|n$). The precise meanings of
these coset spaces should be defined as the orbits of our spinors
$R$. The group in the denominator arises in each case as the
invariance group of the lowest weight state. In the SUSY case,
U($n|n$) corresponds to the superalgebra gl($n|n$) which leaves
the vacuum at each split site invariant, and the notation
indicates that the ordinary group it contains is the compact form,
U($n)\times$ U($n$). The manifold underlying the supermanifold is
thus SO($2n+1$)/U($n$) $\times$ Sp($2n,{\bf R}$)/U($n$), where the
latter factor is non-compact. These (super-) spaces are
homogeneous spaces, but not symmetric (super-)spaces. This implies
that, in a general nonlinear sigma model for these target
manifolds, there is more than one coupling in the kinetic terms
\cite{zinn}---in fact, there are two (see Appendix \ref{betafun}).
The kinetic term is constructed from the metric on the target
manifold, and this metric is not unique, up to a constant factor,
unless the manifold is a symmetric space. Otherwise it is a sum of
two or more pieces. However, as manifolds (without a choice of
metric), these spaces are the same as SO($2n+2$)/U($n+1$) [resp.\
OSp($2n+2|2n$)/U($n+1|n$)] (for the former, this is discussed in
Ref.\ \onlinecite{fulhar}). This is connected with the fact that,
as graded vector spaces, $R$ is the same as the representation
$R_e$. The latter manifolds are symmetric (super-)spaces, and
there is a unique kinetic coupling. The two couplings in the
former point of view are related to the $k$ and $h$ terms. When
$k$ and $h$ on each nearest-neighbor bond are equal, the higher
symmetry implies that the two coupling constants in the kinetic
terms are such that the {\Osp} invariant kinetic term is obtained.
At the same time, $\theta$ can be varied by staggering the
couplings. Counting parameters, there are four in the spin chain
as we have seen, one of which is the overall energy scale which
can be ignored, and also the magnitude of the lowest weight. The
latter controls the magnitude of the kinetic coupling constant.
Ignoring that, the nonlinear sigma model has in general one ratio
of kinetic couplings, and $\theta$, one parameter less than in the
spin chain. However, so far in the nonlinear sigma model we have
assumed Lorentz invariance, whereas in the continuum limit of the
spin chain this need not hold. In fact, in the sigma model, the
same symmetry considerations imply that each of the two pieces of
the kinetic term can be further divided into two terms, which are
the second derivatives in the two orthogonal directions in the 2D
space, and if Lorentz invariance is not required, all the terms
can have different coefficients. That is, a different velocity can
apply to the two kinds of spin waves in the spin chain
\cite{subo(4)}; the two kinds of spin waves correspond to the
decomposition of the small fluctuations around the perturbative
vacuum into two irreducible representations of the symmetry group
U($n$) [or U($n|n$)] (see Appendix \ref{betafun}). Since the
overall energy scale, or one velocity, is a redundant parameter,
this leaves one additional parameter, the ratio of velocities, as
required. Note, however, that if we instead consider the continuum
limit of an isotropic network model, such as the RBIM, only one
velocity can occur. Therefore we would expect that the
universality classes of the transitions would be isotropic and
have a unique velocity.

\begin{figure}
\epsfxsize=3.4in \centerline{\epsffile{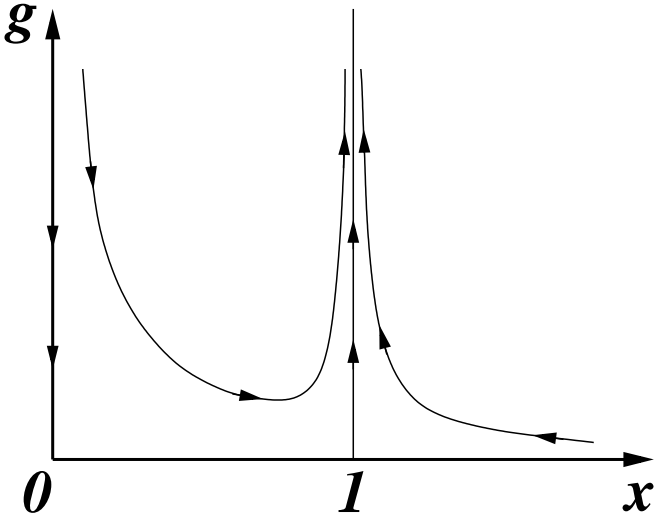}}
\vspace{0.25in} \caption{Sketch of perturbative renormalization
group flows for the couplings $g$, $x$, including two-loop
effects, for the nonlinear sigma model with target space
OSp($2n+1|2n$)/U($n|n$), including a typical flow line for
$g\neq0$, $x\neq0$, $1$.} \label{rgsm}
\end{figure}

Next, it is natural to raise the question of the RG flow of the
two couplings in the {\osp}-invariant nonlinear sigma models. At
least in perturbation theory, the corresponding beta functions
that describe such a flow, are the same as those for the models
with ordinary target manifolds SO($2n+1$)/U($n$), with $n\to 0$.
In Appendix \ref{betafun}, we have computed the beta functions to
one-loop order, neglecting the possibility of more than one
velocity (in some analogous situations, the ratio of velocities
has been shown to renormalize to one \cite{subo(4)}, and we expect
the same to occur here, as also argued in the last paragraph). The
metric on the target manifold is parametrized by two parameters,
say $\eta_1$, $\eta_2$, which appear linearly in the metric, and
the inverses of these parameters are the couplings which are small
at weak coupling, where perturbation theory is valid. The
perturbation expansion has $1/\eta_1$, $1/\eta_2$ for each
propagator, $\eta_1$ or $\eta_2$ for each interaction vertex, so
naively one ends up with a perturbation expansion in, say,
$1/\eta_1$, with each term containing powers, positive or
negative, of the ratio of $\eta_1$ to $\eta_2$. In fact, only
non-negative powers of this ratio appear in the beta functions;
see App.\ \ref{betafun}. The net power (counted with signs) of
$\eta_1^{-1}$'s and $\eta_2^{-1}$'s corresponds to the number of
loops in the Feynman diagrams, as usual. For convenience we will
use the parameters \beq \eta_1 ={1\over g}, \qquad \eta_2 =
{1\over 2xg}. \eeq In this parametrization, $x=1$ is the point
with {\Osp} SUSY (or so($2n+2$) in the replica version). Because
of the higher symmetry at $x=1$, that line should flow onto itself
under RG. Then our one-loop result for the RG flows is ($l$ is the
logarithm of the length scale, as usual)\begin{eqnarray} {d g
\over d l} &=& 2g^2 (x^2-1) + {\cal O}(g^3),\\ { dx \over dl} &=&
- 2 g  x(x-1) + {\cal O}(g^2). \label{betafunctionNzero}
\end{eqnarray}
At $x=1$, the one-loop result is zero, so this line is a line of
fixed points, to this order. This agrees with the one-loop result
for the {\Osp}-invariant model; the beta function to two-loop
order, obtained as the $n\to 1$ limit of that for the
SO($2n$)/U($n$) model (see Eq.\ (\ref{hikbeta})), is \beq {dg
\over dl}=4g^3 + {\cal O}(g^4). \label{hikbetanzero} \eeq Thus, it
vanishes to one-loop order, but not to two-loop order. At two-loop
order, $g$ flows towards strong coupling (see Fig.\ \ref{rgsm}).
The one-loop flows for $x\neq 1$ take $x$ closer to 1, and (except
for $x=0$) the flows starting at $g\neq 0$ never reach $g=0$.
Instead they flow to the region $x\simeq 1$ where the one-loop
terms vanish, and the two-loop term cannot be neglected. Since the
two-loop term in $dg/dl$ at $x=1$ is positive, all flows from weak
coupling eventually go towards large $g$, with $x$ approaching 1,
{\em except} when $x=0$. On the latter line, $g$ flows towards
weak coupling. As discussed in App.\ \ref{betafun} (but here in
SUSY language), on this line the larger SUSY has spontaneously
broken, and the system is described by the OSp($2n|2n$)/U($n|n$)
nonlinear sigma model [there is an additional global degree of
freedom, described by a point on OSp($2n+1|2n$)/OSp($2n|2n$), a
``supersphere'', on which the larger SUSY algebra acts]. Flows
that begin at small, nonzero $x$ eventually go to strong coupling.
This generates very large crossover lengths, due to the very slow
flows near $x=1$, $g=0$, where the first nonzero term is at
two-loop order; the length scale at which $g$ becomes of order one
is of order $\exp[1/(2g_0x_0)+1/(8g_0^2x_0^2)]$, in units of a
short distance cutoff, where $g_0$ and $x_0$ are the bare values
of $g$ and $x$, and it was assumed that $x_0$ and $g_0x_0$ are
both small (see App.\ \ref{betafun}).

As usual, the perturbative results do not depend on $\theta$, but
such dependence can be expected nonperturbatively. For the sigma
model with {\Osp} SUSY, and for $x\neq0$, the flows go towards
strong coupling, and it is highly plausible, based on our
experience with transitions (such as the integer quantum Hall
transition) with such behavior of the couplings, that there is a
unique fixed point at strong coupling $g$ and at $\theta=\pi$ (mod
$2\pi$). Hence we expect that the spin chain at point $O$ has the
same critical theory as the {\Osp}-invariant sigma model. It is
also quite plausible, based on the behavior of the flows, that at
least models that map onto the nonlinear sigma models with {\osp}
SUSY also flow to the same critical theory, when $\theta=\pi$ (mod
$2\pi$). It is still possible that in our generalized spin chains,
points other than $O$ do not flow to the {\Osp} critical theory,
but it is plausible that there is a nontrivial neighborhood of $O$
on the critical surface that does. This possibility may seem more
plausible if we point out that in some other cases (without SUSY),
a similar phenomenon is believed to occur \cite{subo(4)}. It is of
course less clear that distant points such as $N$, $K$, and $H$
flow to the same theory. Since the spin chain models typically
start at bare couplings of order 1, we can almost rule out any
flow to the weak-coupling regime of the {\osp} or {\Osp}-invariant
nonlinear sigma models, (analogous to that in the lower-SUSY {\os}
nonlinear sigma model \cite{bundschuh,sfnew,readgr,bsz}) because
that regime is not stable under the RG. A flow to weak coupling is
only possible by tuning a parameter, corresponding to putting
$x=0$ or $g=0$ in the weak-coupling analysis. A natural guess is
that the $h$-only models might satisfy one of these conditions.
This might even occur for a range of values of the staggering,
corresponding to changing $\theta$ away from $\pi$ (mod $2\pi$) in
the nonlinear sigma model, since the value of $\theta$ is
irrelevant (or formally, exactly marginal) at weak coupling. We
have been unable to see why any of these models should satisfy
such a condition exactly. However, it may be that one of them lies
close to $x=0$. In that case, the RG flows take them close to
$g=0$, and since the flows to strong coupling pass near $x=1$,
$g=0$, where the first nonzero term is at two loops, the crossover
length could be very large. That is, very large systems would be
needed to see the true asymptotic critical behavior. Another
possibility is that these models lie at bare values $x>1$, $g$
very small, which again yields a large crossover length. Since we
have a two-parameter space of critical models, we may expect to be
able to tune $g$ or $x$ small somewhere in this space. However,
arguments presented elsewhere \cite{rl} show that the RBIM, and
hence the point $N$, cannot flow to the weak coupling region.

The leading alternative scenario seems to be that although most
points in the phase diagram flow to the {\Osp}-invariant fixed
point, the multicritical point $N$ on the true NL,  may be a
distinct universality class, and the perturbation off this point
in the phase diagram may be a relevant one that causes a flow to
the {\Osp}-invariant fixed point. Clearly, we cannot answer here
the question of which of these scenarios is correct. But the
self-duality apparent in the {\Osp}-invariant model at its
critical point, as manifested by the reflection symmetry about a
split site in the chain, and the significant lack of it in the
RBIM, which instead has the special ``symmetry'' property that the
Ising couplings are all real \cite{rl}, suggests that this
alternative scenario may be correct.

There is one further point to make about the spin chains and
nonlinear sigma models, that applies {\em off} the NL. In that
case the SUSY is broken to {\os}. It will be convenient here to
make use of the replica formalism, in which the language and
notation are simpler and standard, but the ideas extend also to
the supergroups, since the additional bosons and Sp($2n,{\bf R}$)
symmetry, and the odd generators, do not change the form of the
argument. In the higher SUSY Hamiltonians, including that for the
NL, the global symmetry group can be seen to be SO($2n+1$), and
making any of the $\lambda$ parameters $\neq 1$ reduces the
symmetry group to O($2n$), not just SO($2n$) [there seems to be no
accepted notation in the supergroup case for the distinction
analogous to that between O($N$) and SO($N$), nor for that between
SO($N$) and its covering group Spin($N$), which we ignore here].
Furthermore, the representations $R$, $\bar{R}$ are reducible
under SO($2n$); they split into two nonisomorphic irreducible
spinors, each of dimension $2^{n-1}$. These two spinors correspond
to even and odd numbers of fermions in the unconstrained
representation (see App.\ \ref{reposp}). However, under O($2n$),
$R$, $\bar R$ do not split; O($2n$) has irreducible
representations of dimension $2^n$ [this is related to the fact
that O($2n$) is not a direct product of SO($2n$) with ${\bf Z}_2$,
unlike the case of O($2n+1$)]. Thus we will still call the models
spin chains, since they involve irreducible representations of
their symmetry group, O($2n$).

When we consider the corresponding nonlinear sigma models, via the
usual correspondence, we naturally consider the orbit of the
lowest weight in $R$ under O($2n$). Due to the disconnected nature
of O($2n$), as opposed to SO($2n$), this orbit O($2n$)/U($n$)
falls into two disconnected pieces, which are both of the form
SO($2n$)/U($n$) as manifolds. Similar statements hold for the
supermanifolds in the SUSY formalism.

In a recent paper \cite{bsz} on the class D of random matrix
problems, which is the same symmetry class as the RBIM fermion
problem we are considering, it was emphasized that the target
manifold of the nonlinear sigma model that describes it is
O($2n$)/U($n$) in the replica formalism, which has two connected
components, corresponding to those of the group O($2n$) [or the
corresponding supergroup OSp($2n|2n$)]. This opens a possibility
not usually considered for nonlinear sigma models, that the
configurations include fluctuations (i.e.\ domains) where the
sigma model field is on different components of the target
manifold. This implies that additional parameters, beyond the
usual couplings like $g$, $\theta$ for continuous deformations of
the field, must appear in the model, to describe the domain walls;
for example, a fugacity per unit length of domain wall. When the
fugacity is small, there are essentially no domain walls, and the
model would reduce to that with target space SO($2n$)/U($n$).

In our approach, we have arrived at spaces of states that
correspond to both parts of the target manifold, and further the
spin chain Hamiltonians contain in general eight parameters.
Therefore, our spin chain models describe a strong-coupling
version of the physics of the nonlinear sigma model with domain
walls included. These models include the pure Ising model, and
weak-disorder, limits. Note that the latter are not accessible
simply as the strong coupling, $g\to\infty$, limit of the
SO($2n$)/U($n$) nonlinear sigma model (compare Ref.\
\onlinecite{sfnew}).

What we have found in this paper is that the states in the SUSY
description, can be viewed not only as domains of two ``phases'',
but that the discrete (Ising-like) degree of freedom, which labels
which phase [component of the target manifold, or irreducible
spinor of SO($2n$)] a point in 2D space is in, can be replaced by
additional continuous variables. These continuous degrees of
freedom turn the model into a nonlinear sigma model with symmetry
SO($2n+1$) [or SUSY {\osp}] broken by certain terms in the action,
or else a strong-coupling version of this, at least near the NL.
This may be of future use in uncovering the physics of these
general class D problems, not only the RBIM. The replicated spin
chains for O($2n$) at nonzero $n$ have not been considered
previously (except for the $n=1$ case, the usual XXZ model), and
are also of interest in their own right. Note finally that in our
earlier discussion of SO($2n+1$)- and SO($2n+2$)-invariant models,
the representations of the stated groups were irreducible, the
corresponding target manifolds were connected, and no analogous
domain walls were possible.

\section{Conclusion}
\label{conclusion}

In this paper we applied the supersymmetry (SUSY) method to
analyze an Ising model with a binary distribution of random bonds
(RBIM). The Nishimori line (NL) on the phase diagram of the model
is a line with the enhanced SUSY \osp. On the rest of the phase
diagram the model has only {\os} SUSY. The enhanced SUSY on the
Nishimori line allows us to rederive the identities (\ref{correq})
among various correlation functions. More generally, we have shown
that the transition on the NL has very strong analogies with the
integer quantum Hall effect transition, and other random fermion
problems in 2D, such as the spin quantum Hall transition, which
can also be modeled by (super-)spin chains with alternating dual
irreducible representations at the sites, and staggered couplings.
The conformal field theories of the critical points are mostly
unknown at present. We emphasize that, in view of our results and
those of Ref.\ \onlinecite{ldgh}, the fixed-point conformal field
theory of the multicritical point in the RBIM with a {\em generic}
distribution for the bonds (not only those satisfying the
Nishimori condition) must have at least {\osp} SUSY, and this is a
requirement for any future proposal for a conformal field theory
of the multicritical point within the SUSY formulation. We have
also demonstrated that such higher SUSY points occur in other
problems, such as a 1D model, and probably elsewhere. After
analyzing the phase diagram of generalized Hamiltonians with the
same enhanced SUSY as the NL, we suggested that the transitions in
many or all of these more general 2D models are in a universality
class with a still larger SUSY, {\Osp}. This hypothesis is
supported to some extent by the weak-coupling RG analysis of the
nonlinear sigma models that correspond to the spin chains.

Fitting our results into the framework of random matrix ensembles
for such problems is an outstanding challenge. It is interesting
that the nonlinear-sigma--model target manifold we obtain on the
NL is (except for $x=0$) {\em not} in the list of those known to
correspond to random matrix ensembles in Ref. \onlinecite{az}.
Possibly there is another random matrix theory with special
symmetries as on the NL.

There are of course a number of other outstanding problems, even
for the RBIM. We have hardly touched the region below the NL,
which remains mysterious. The fixed point at $K=\infty$ (zero
temperature) and $p=p_c$ is of particular interest. In this region
the system can be viewed as a superspin chain, since it is a chain
of irreducible representations of its supergroup, OSp($2n|2n$), to
which the larger SUSY, OSp($2n+1|2n$), is broken by superspin
anisotropy terms, similar to the XXZ model.

{\it Note added:} Another numerical work on the multicritical
point of the $\pm K$ RBIM has appeared very recently \cite{hpp}.

\acknowledgments

We thank S. Sachdev for pointing out Ref.\ \onlinecite{subo(4)}.
This work was supported by NSF grants, Nos.\ DMR--91--57484 (IAG
and NR), DMR--98--18259 (NR). The research of IAG and NR was also
supported in part by NSF grant No.\ PHY94--07194. AWWL was
supported in part by the A. P. Sloan Foundation.


\appendix

\section{Equalities for correlators}
\label{eqcorr}

We will show in this Appendix that the enhanced supersymmetry
present on the Nishimori line in our formulation allows us to
reproduce the results of the type of Eq.\ (\ref{correq}).

We use the formulation of the correlators in the Ising model in
terms of paths and modified partition functions \cite{kc,id}.
Namely, for a correlator of two spins $S_{\bi_1}$ and $S_{\bi_2}$
we join the points $\bi_1$ and $\bi_2$ by an (arbitrary) path on
the lattice, shift all the coupling constants $K$ by $i\pi/2$
along the path, and calculate the modified partition function
$Z^{\text{(mod)}}$ for the system with the modified couplings.
Then the correlator is \beq \la S_{\bi_1} S_{\bi_2} \ra = (-i)^l
\frac{Z^{\text{(mod)}}}{Z}, \label{corr} \eeq where $l$ is the
length of the path.

In the quantum formalism the vertical coordinate on the original
square lattice plays the role of imaginary time $\tau$, and the
partition function is given by the supertrace of an imaginary time
ordered evolution operator $U$, composed of the transfer matrices
$T_{h\bi}$ and $T_{v\bi}$ for all the bonds in the model. Because
of the supersymmetry the partition function equals 1 by
construction (see Eq.\ (\ref{susycond}) and following) for any
realization of the random couplings: \beq Z_{\rm SUSY} = \STr \Tt
U = 1. \eeq When calculating the correlator (\ref{corr}) we have
to modify the couplings in the transfer matrices along the path
only for one particular replica, say, the first  fermionic one.
Then the correlator will be \beq \la S_{\bi_1} S_{\bi_2} \ra =
(-i)^l \STr \Tt U^{\text{(mod)}}. \eeq Similarly, when calculating
$\la S_{\bi_1} S_{\bi_2} \ra ^{2m-1}$, we have to modify the
couplings for $2m-1$ different replicas, in which case we must
have $2n > 2m-1$.

Let us see how the transfer matrices are modified, when we
shift the couplings by $i\pi/2$. Start with the horizontal
transfer matrix, assuming that the couplings are modified for
$2m-1$ fermionic replicas:
\bea
T_{h\bi}^{\text{(mod)}} &=&
\exp \left(2 K_{\bi,\bi + \bx} \XSi + i\frac{\pi}{2}\sum_{\al = 1}^{2m-1}
2 x_{\al i}\right) \no\\
&=& i^{2m-1} \, T_{h\bi} \prod_{\al = 1}^{2m-1} (2 x_{\al i}).
\eea
Upon averaging over the randomness this becomes
\beq
T_{2i}^{\text{(mod)}} = i^{2m-1} \, T_{2i}
\prod_{\al = 1}^{2m-1} (2 x_{\al i}).
\eeq
Then for the correlator of two spins in the same horizontal row
we have
\beq
\disave{\la S_{\bi} S_{\bi+r\bx} \ra^{2m-1}} =
\STr \Tt V \prod_{k = i}^{i+r-1}
\prod_{\al = 1}^{2m-1} (2 x_{\al k}),
\eeq
where $V = \disave{U}$.

As before for single transfer matrices, we can rewrite the last
expression in terms of operators, acting in the space $\cal F'$,
using the substitution rules obtained above in Sec.\
\ref{enhsusy}: \beq \disave{\la S_{\bi} S_{\bi+r\bx} \ra^{2m-1}} =
\STr \Tt V' \prod_{k = i}^{i+r-1} (2 x_{0k}) \prod_{\al =
1}^{2m-1} (2 x_{\al k}). \eeq Now comes the crucial point. On the
Nishimori line the zeroth fermion is supersymmetric with the rest
of the replicas, so we can replace all $x_{0k}$ in the last
expression by, say, $x_{2m, k}$: \beq \disave{\la S_{\bi}
S_{\bi+r\bx} \ra^{2m-1}} = \STr \Tt V' \prod_{k = i}^{i+r-1}
\prod_{\al = 1}^{2m} (2 x_{\al k}). \eeq Then we can safely go
back to the original space $\cal F$, in which the last expression
is easily identified as \beq \disave{\la S_{\bi} S_{\bi+r\bx}
\ra^{2m}}, \eeq which proves the relation (\ref{correq}) for this
particular case.

Now see how vertical transfer matrices are modified. When we
modify the coupling for the fermionic replica 1 on a vertical
bond, the vertical transfer matrix for this replica is modified
from \beq \frac{e^{K_{\bi,\bi+\by}} e^{\tilde
K_{\bi,\bi+\by}}}{\cosh \tilde K_{\bi,\bi+\by}} \exp\left(-2
\tilde K_{\bi,\bi+\by} n_{1i} \right) \eeq to \bea &&i
\frac{e^{K_{\bi,\bi+\by}} e^{-\tilde K_{\bi,\bi+\by}}}{\cosh
\tilde K_{\bi,\bi+\by}} \exp\left(2 \tilde K_{\bi,\bi+\by} n_{1i}
\right) \br = 2i \sinh K_{\bi,\bi+\by} \exp\left(2 \tilde
K_{\bi,\bi+\by} n_{1i} \right). \eea (since when $K$ is shifted by
$i\pi/2$, the dual coupling $\tilde K$ changes sign). Adding the
rest of fermionic and bosonic replicas, we obtain
\beq
T_{v\bi}^{\text{(mod)}} = i (\tanh K_{\bi,\bi+\by})^{\NSi -
2n_{1i} +1}. \eeq
If we modify the coupling for replicas 1 through
$k$, we get similarly \beq T_{v\bi}^{\text{(mod)}} = i^k (\tanh
K_{\bi,\bi+\by})^{\NSi - \sum_{\al =1}^{k} (2n_{\al i} - 1)}. \eeq

To average this expression, we have to distinguish the cases of
odd and even $k$. For an even $k = 2m$ we get \bea
T_{1i}^{\text{(mod)}} &=& i^{2m} (\tanh K)^{\NSi - \sum_{\al
=1}^{2m} (2n_{\al i} - 1)} \br \times  \left(1 - p +
p(-1)^{\NSi}\right) \no\\ &=& i^{2m} T_{1i} \prod_{\al =1}^{2m}
y_{\al i}, \eea where we introduced \beq y_{\al i} = e^{2K^*
(2n_{\al i} - 1)}. \eeq The corresponding operator in $\cal F'$ is
\beq T_{1i}^{\prime {\text{(mod)}}} = i^{2m} T_{1i}' \prod_{\al
=1}^{2m} y_{\al i}. \eeq Then for the correlator of two spins in
the same column we get \bea \disave{\la S_{\bi} S_{\bi+r\by}
\ra^{2m}} &=& \STr \Tt V \prod_{\tau = j}^{j+r-1} \prod_{\al =
1}^{2m} y_{\al i}(\tau) \no\\ &=& \STr \Tt V' \prod_{\tau =
j}^{j+r-1} \prod_{\al = 1}^{2m} y_{\al i}(\tau). \label{vertcorr}
\eea

For odd number $k = 2m-1$ we obtain instead upon averaging \bea
T_{1i}^{\text{(mod)}} &=& i^{2m-1} (\tanh K)^{\NSi -
\sum_{\al=1}^{2m-1} (2n_{\al i} - 1)} \br \times \left(1 - p +
p(-1)^{\NSi + 1}\right). \label{t1mod} \eea The last factor here
is different from the similar factor in $T_{1i}$. Now it gives 1
when $\NSi$ is odd, and $1-2p = e^{-2L^*}$ when $\NSi$ is even. In
the space ${\cal F}'_i$ this factor may be written as
$e^{2L^*(n_{0i} - 1)}$, and the Eq.\ (\ref{t1mod}) is replaced by
\beq T_{1i}^{\prime {\text{(mod)}}} = i^{2m-1} T_{1i}' \, y_{0i}
\prod_{\al =1}^{2m-1} y_{\al i}, \eeq where \beq y_{0i} =
e^{2L^*(2n_{0i} - 1)}. \eeq For the vertical correlator we obtain
now \beq \disave{\la S_{\bi} S_{\bi+r\by} \ra^{2m-1}} = \STr \Tt
V' \prod_{\tau = j}^{j+r-1} y_{0i}(\tau) \prod_{\al = 1}^{2m-1}
y_{\al i}(\tau). \label{vertcorr1} \eeq On the Nishimori line due
to the enhanced supersymmetry (and the fact that $L^* = K^*$) the
factor $y_{0i}$ may be replaced by $y_{2m, i}$ and we again get
the equality of the type of Eq.\ (\ref{correq}).

The structure appearing in the formulation above for the correlators is
multiplicative in the bonds which are modified along a path connecting the
spins. Then it is straightforward to generalize the arguments of this
appendix to the case of arbitrary spin correlators.

\section{Representations $R$ and $\bar R$}
\label{reposp}

In this Appendix we review the construction of the representations
$R$ and $\bar R$ of osp($3|2$) in terms of {\it unconstrained\/}
fermions and bosons (for details see Ref.\ \onlinecite{fss}). We
also discuss how to form a graded tensor product of such
representations and obtain the invariant product of the superspins
$G$ and ${\bar G}$.

To construct the representation $R$ we need only one complex boson
$b$ and one complex fermion $f$ and their conjugates $\bd$, $\fd$, with
usual commutation relations. In terms of these the generators of
osp($3|2$) are constructed as follows. For an orthonormal basis of the
Cartan subalgebra we use
\beq
h_1 = {1 \over \sqrt 2}\left(\bd b + {1\over 2}\right), \quad
h_2 = {i \over \sqrt 2}\left(\fd f - {1\over 2}\right).
\label{r1}
\eeq
In the distinguished system of simple roots of osp($3|2$) one root $\al_1$
is odd (``fermionic''), and one root $\al _2$ is even (``bosonic'').  The
generators corresponding to these roots (and their negatives) are
\bea
e_{\al_1} &=& \bd f, \quad e_{-\al_1} = \fd b, \no\\
e_{\al_2} &=& (-1)^{\nf} \fd, \quad
e_{-\al_2} = f (-1)^{\nf}.
\label{r2}
\eea
The other roots are $\al_3 = \al_1 + \al_2$, $\al_4 = \al_1 + 2
\al_2$ (both odd), $\al_5 = 2 \al_1 + 2 \al_2$ (even), and
their negatives. The corresponding generators are
\bea
e_{\al_3} &=& (-1)^{\nf} \bd, \quad
e_{-\al_3} = b (-1)^{\nf},
\no\\
e_{\al_4} &=& \bd \fd, \quad e_{-\al_4} = f b, \no\\
e_{\al_5} &=& (\bd)^2, \quad e_{-\al_5} = b^2.
\label{r3}
\eea

Note that the generators corresponding to the roots $\al_2$ and
$\al_3$ contain expression $(-1)^{\nf}$. This is a ``twist''
operator for the fermion, which means that it anticommutes with
$f$ and $\fd$. It is necessary to ensure that these generators
obey the (anti-)commutation relations. In other words, these
choices reflect the grading appropriate for osp($3|2$), instead of
that which is natural in the present Fock space.

\begin{figure}
\epsfxsize=3.4in \centerline{\epsffile{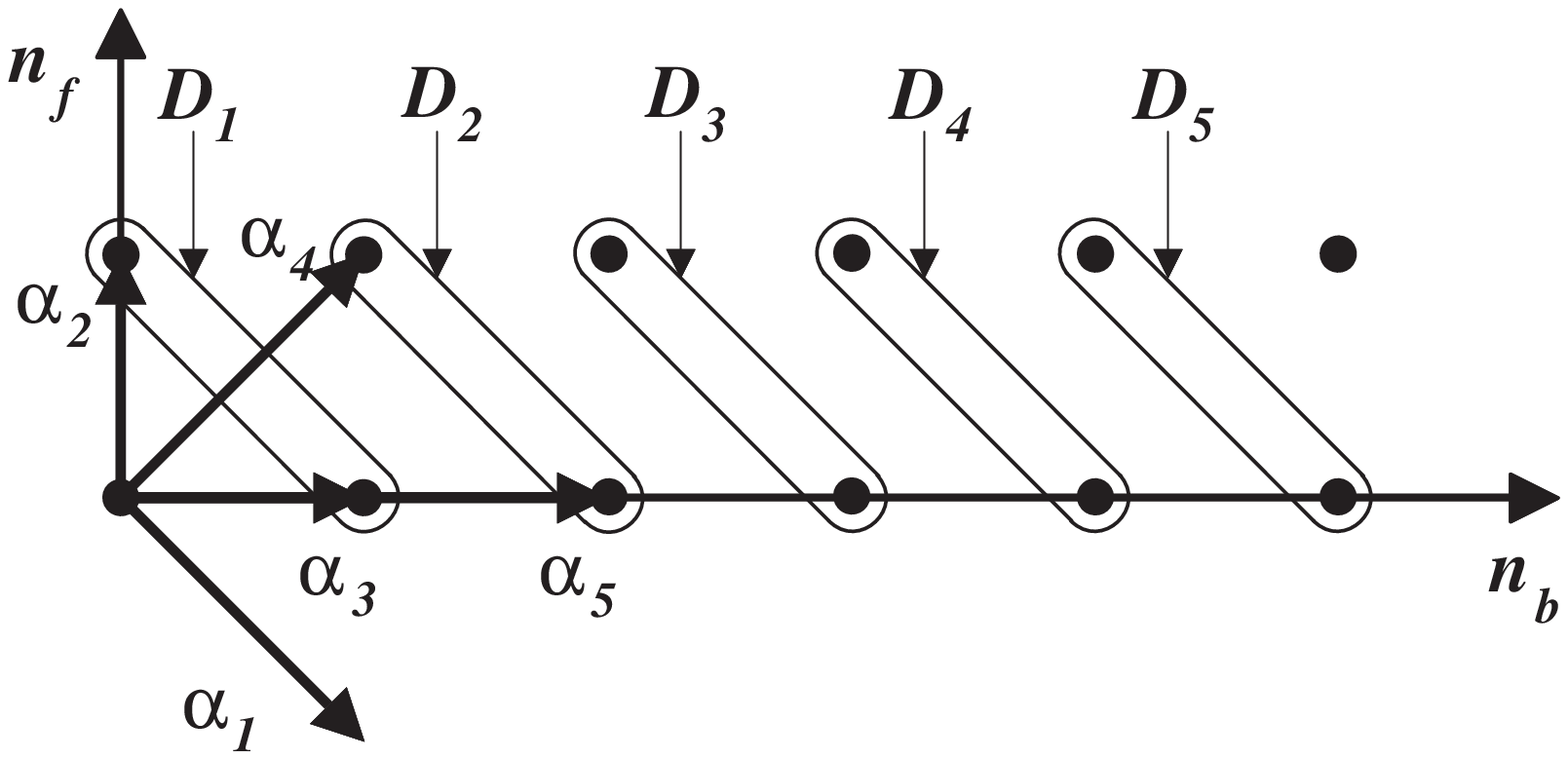}}
\vspace{0.25in} \caption{The weights of states in $R$. The action
of the positive root generators is shown by arrows. The states are
grouped in pairs which are the doublets under gl($1|1$) (see text
for details).} \label{fig7}
\end{figure}

The vacuum for bosons and fermions $|0 \ra$, defined in the usual
manner \beq b|0\ra = f|0\ra =0 \eeq is the lowest weight state of
the $R$ representation. The remaining states are obtained by the
action of the raising generators, and it is easy to see that they
span the whole Fock space of $b$ and $f$. The weights of the
states in $R$ in terms of $n_f$ and $n_b$ are shown in Fig.\
\ref{fig7}. We also show in this Figure the organization of the
states in doublets under the gl($1|1$) subalgebra generated by
\bea E &\equiv& \nb + \nf, \quad N \equiv \frac{1}{2}(\nb - \nf),
\no\\ F^{\dagger} &\equiv& e_{\al_1} = \bd f, \quad F \equiv
e_{-\al_1} = \fd b, \eea (see Ref. \onlinecite{rs}, which contains
detailed discussion of the irreducible representations of
gl($1|1$).) The doublet of states with $E=m$ is denoted by $D_m$.

{}From Fig.\ \ref{fig7}, we can see that the grading of states,
consistent with that of the SUSY generators, and such that the
vacuum (lowest weight state) is even, is that states are even or
odd according as the number of {\em bosons} is even or odd. This
agrees with the choice we made in Sec.\ \ref{tmatr} for other
reasons. We may also note that the generators without strings,
which are bilinears in the bosons and fermions, generate the
osp($2|2$) subalgebra. The latter algebra is consistent with the
natural grading on the Fock space. This is not in contradiction to
the above construction, because the Fock space decomposes into two
irreducible ``spinor'' representations of osp($2|2$), which are
connected to each other only by the shortest roots $e_{\pm 2}$,
$e_{\pm 3}$ that are not present in osp($2|2$).

The construction of $\bar R$ is similar. The difference is that we
start with negative norm bosons $\bb$ and $\bbd$ satisfying \beq
[\bb, \bbd] = -1, \label{com} \eeq and another pair of the usual
fermionic operators $\fb$ and $\fbd$. One possible choice of the
generators of osp($3|2$) in the $\bar R$ representation is \bea
\hb_1 &=& {1 \over \sqrt 2}\left(\bb\bbd + {1\over 2}\right),
\quad \hb_2 = {i \over \sqrt 2}\left(\fb\fbd - {1\over 2}\right),
\no\\ \eb_{\al_1} &=& \bb \fbd, \quad \eb_{-\al_1} = \fb \, \bbd,
\no\\ \eb_{\al_2} &=& -(-1)^{\nbb} \fb, \quad \eb_{-\al_2} = -\fbd
(-1)^{\nbb}, \no\\ \eb_{\al_3} &=& \bb (-1)^{\nbb}, \quad
\eb_{-\al_3} = (-1)^{\nbb} \bbd, \no\\ \eb_{\al_4} &=& -\bb \fb,
\quad \eb_{-\al_4} = -\fbd \bbd, \no\\ \eb_{\al_5} &=& -\bb^2,
\quad \eb_{-\al_5} = -(\bbd)^2, \eea Here the number of $\bb$
bosons is defined as \beq \nbb = - \bbd\bb. \eeq The minus sign in
this expression implies that $\nbb$ is a nonnegative integer, with
eigenstates $|\nbb \ra = (\nbb!)^{-1/2} (\bbd)^{\nbb} |{\bar
0}\ra$. Note that now the $\al_2$ and $\al_3$ generators contain a
twist operator for the boson $\bb$, which also ensures the proper
(anti-)commutators.

Now the vacuum $|{\bar 0}\ra$ for $\fb$ and $\bb$ defined as
\beq
\fb|{\bar 0}\ra = \bb|{\bar 0}\ra = 0
\eeq
is the highest weight state of the $\bar R$ representation, and the
remaining states span the whole Fock space of $\fb$ and $\bb$. The
states of $\bar R$ are now organized in doublets ${\bar D}_{\bar m}$ of
the gl($1|1$) generated by
\bea
\bar E &\equiv& -\nbb - \nfb,
\quad \bar N \equiv \frac{1}{2}(\nfb - \nbb), \no\\
{\bar F}^{\dagger} &\equiv& \eb_{\al_1} = \bb \fbd,
\quad \bar F \equiv \eb_{-\al_1} = \fb \, \bbd.
\eea

Next we have to combine the representations $R$ and $\bar R$ in
the alternating fashion, as in Fig.\ \ref{fig2}. When we try to do
that we immediately realize that the twist operators of individual
$R$ and $\bar R$ representations are not adequate for their job in
the tensor product. They should be replaced by ``strings'',
similar to the ones used in the Jordan-Wigner transformation. One
possible convenient choice of these strings is the following. For
the representation $R_i$ (numbering as in the Fig.\ \ref{fig2})
the twist operator $(-1)^{n_{fi}}$ is replaced by \beq \Si_i =
\prod_{k \leq i} (-1)^{n_{fk} + n_{\fb k}} \prod_{k > i}
(-1)^{n_{bk} + n_{\bb k}}. \eeq Similarly, for ${\bar R}_i$ the
operator $(-1)^{n_{\bb i}}$ should be replaced by \beq {\bar
\Si}_i = \prod_{k < i} (-1)^{n_{fk} + n_{\fb k}} \prod_{k \geq i}
(-1)^{n_{bk} + n_{\bb k}}. \eeq Note that in fact, $\Si_i = {\bar
\Si}_{i+1}$.

For the purposes of Sec.\ \ref{lowtemp} we need to consider only
one pair of the antiferromagnetically coupled superspins. In this
case the $\Si$ operator common for both representations is $\Si =
(-1)^{\nf + \nbb}$. First we consider the fully osp($3|2$)
invariant product appearing in $H_k$ on the NL. In terms of the
root generators this product is given by \beq \str G{\bar G} =
{}-h_1 \hb_1 - h_2 \hb_2 - \sum_{\al > 0} K_{- \al}^{-1}
\left(e_\al \eb_{-\al} + \eb_\al e_{-\al} \right), \eeq where
$K_{-\al} = K(e_{-\al}, e_\al)$ are the values of the Killing form
on the pairs $e_{-\al}, e_\al$. For osp($3|2$)  they are \bea
K_{-\al_1} &=& -2, \quad K_{-\al_2} = -4, \quad K_{-\al_3} = -4,
\no\\ K_{-\al_4} &=& -2, \quad K_{-\al_5} = -4. \eea With the
mentioned expression for $\Si$ this gives \bea -4 \str G{\bar
G}&=& (\bd)^2(\bbd)^2 + \bb^2 b^2 + 2(\bd \bbd + \bb b)(\fd \fbd +
\fb f) \br - 2\nb\nbb - \nb - \nbb + 2\nf\nfb - \nf - \nfb \br
-(\fd\fbd + \fb f + \bd\bbd + \bb b) \no\\ &=& J^2 - J, \label{gg}
\eea where \beq J = \fd\fbd + \fb f + \bd\bbd + \bb b. \eeq

Note that the term $J$ in Eq.\ (\ref{gg}) comes from the roots
$\pm \al_2$ and $\pm \al_3$, and $J^2$ comes from all the
remaining roots. These remaining roots are exactly the roots of
osp($2|2$). Therefore, the $J^2$ term is the osp($2|2$) invariant
product. That observation allows us to write the general
anisotropic product as \beq 4\str \La G \La {\bar G} = \lam J-J^2.
\eeq

\section{Perturbative beta function for
SO($\lowercase{2n+1}$)/U($\lowercase{n}$)}
\label{betafun}

In this Appendix we derive the perturbative beta function of the
weakly-coupled nonlinear sigma model on SO($2n+1$)/U($n$) target
space, to one-loop order. The underlying ideas for a general sigma
model have been discussed extensively  by Friedan \cite{Friedan},
and we can be brief (similar calculations can be found in Ref.\
\onlinecite{subo(4)}). We then discuss the resulting flows at
$n=0$.

Consider a general homogeneous space $G/H$, where $H$ is a
subgroup of the group $G$. The neighborhood of points $g H \in
G/H$ of the ``origin'' $O\equiv e H$ ($g$ is an arbitrary element,
and $e$ is the identity, in $G$) may be parametrized in terms of
$\dim G-\dim H$ coordinates $X^I$ by writing $g=\exp\{ X^I T_I \}$
(repeated indices summed). Here $T_I$ denotes a basis of the
vector space ${\cal G}/{\cal H}$ spanned by the generators of $G$
which are not generators of $H$ (${\cal G}$ and ${\cal H}$ denote
the Lie algebras). The sigma model on $G/H$ is then defined by the
action \beq S = {1\over 2} \int d^2 r \, \eta_{IJ} \{ X(r)\}
\partial_\mu X^I(r)
\partial^\mu X^J(r) \label{sigmaaction} \eeq where $r$ is the
coordinate of two-dimensional space. The metric $\eta_{IJ} \{ X
\}$ on the target space $G/H$ of the sigma model serves as the
coupling constant(s). Every point $P=g H$ in the coset space can
be reached from the origin by left multiplication, and every
element of the tangent space at $P$ can be similarly obtained from
the tangent space ${\cal G}/{\cal H}$ at the origin. Therefore,
the metric at any point $P$ is uniquely determined by that at the
origin $O$, where it represents a symmetric  bilinear form on the
vector space ${\cal G}/{\cal H}$. In order for this bilinear form
to represent a metric at the origin it must be invariant under the
subgroup $H$ (which acts by conjugation). Therefore, the metrics
on the homogeneous space are in 1-1 correspondence with
$H$-invariant symmetric bilinear forms on ${\cal G}/{\cal H}$.

For sigma models on general manifolds (not necessarily homogeneous
spaces) the two-loop beta function is \cite{Friedan} \beq { d
\eta_{ij}(X) \over dl} = R_{ij}(X) + {1\over 2} R_{iklm}(X)  \
{R_j}^{klm}(X) +\ldots, \label{betaFriedan} \eeq where $X^i$ is
any system of local coordinates, and ${R^i}_{klm}(X)$, $R_{ij}(X)$
are the Riemann and Ricci tensors, respectively, at the point of
the manifold with coordinates $X$.

For a homogeneous space $G/H$, it is enough to compute the beta
function  for the metric at the origin $O=e H$ (since all other
points can be reached by left multiplication with elements of $G$,
acting  as isometries), where it reads \beq { d \eta_{IJ} \over
dl} = R_{IJ} + {1\over 2} R_{IKLM}  \ {R_J}^{KLM} +\ldots.
\label{betacoset} \eeq This form of the beta function  is
convenient since it does not require reference to any
parametrization of the coset space. Rather, the Riemann tensor of
the homogenous space \cite{Friedan,Kobayashi}, viewed as a
Riemannian space, has a simple expression in terms of of the
structure constants ${f_{IJ}}^K$, ${f_{IJ}}^a$ of the Lie algebra
$\cal G$, and the metric:
\begin{eqnarray}
R_{KLIJ}&=&{} -{1\over 4} \bigl ( f_{IML} - f_{MLI} + f_{LIM}
\bigr )  \eta^{M M'}\nonumber\\ &&{}\qquad{}\times \bigl (
f_{JKM'} - f_{KM'J} + f_{M'JK} \bigr ) \nonumber\\&&{} + {1\over
4} \bigl ( f_{JML} - f_{MLJ} + f_{LJM} \bigr ) \eta^{M M'}
\nonumber\\&&{}\qquad{}\times\bigl ( f_{IKM'} - f_{KM' I} + f_{M'
I K} \bigr ) \nonumber\\&&{} + {1\over 2} {f_{IJ}}^M \ \bigl (
f_{MKL} - f_{KLM} + f_{LMK} \bigr ) \nonumber\\&&{} + {f_{IJ}}^a \
f_{aKL}, \label{Riemann}
 \end{eqnarray}
where indices $K, L, M, M'$ denote generators in ${\cal G}/{\cal
H}$, which are lowered and raised by means of the metric
$\eta_{IJ}$ and its inverse $\eta^{IJ}$. Indices $a$ denote
generators in $\cal H$. The Ricci tensor is obtained, as usual, by
contraction,
\begin{equation}
R_{LJ}
=
\eta^{KI} \ R_{KLIJ}. \label{Ricci} \end{equation}

We now discuss the space of all possible ($G$-invariant) metrics
on the homogeneous space, that is  all $H$-invariant symmetric
bilinear forms on the vector space ${\cal G}/{\cal H}$. Since the
latter transforms in a (real) representation of $H$ (under
conjugation), the bilinear form $\eta_{IJ}= \eta( T_I, T_J)$ must,
by Schur's lemma, be a multiple of the unit matrix on each
irreducible component (assuming, for simplicity, that each such
component occurs only once). Consider, for example, the
homogeneous spaces SO($N$)/SO($N-1$), the familiar O($N$) vector
models. Here ${\cal G}/{\cal H}=$ so($N$)/so($N-1$) transforms in
the (irreducible) vector representation of SO($N-1$) and therefore
there is only a one-parameter family of metrics. This is the case
for all {\em symmetric} spaces \cite{helgason}, whose sigma models
have therefore only a single coupling constant (the scale of the
metric).

The case of interest in this paper is $G/H=$ SO($2n+1$)/U($n$),
which is not a symmetric space. It has a {\em two}-parameter
family of metrics, and the corresponding sigma model has therefore
two coupling constants. To see this one notes that the vector
space so($2n+1$)/u($n$) decomposes (over the real numbers) under
the adjoint action of U($n$) into two irreducible representations.
One of them is of dimension $n(n-1)$; the corresponding generators
will be denoted $T_{I_1}$. The other is of dimension $2n$, and the
corresponding generators will be denoted $T_{I_2}$. These vector
spaces may be identified with the cosets of Lie algebras
so($2n$)/u($n$) and so($2n+1$)/so($2n$), respectively. This
decomposition corresponds to the chain of subalgebras, u($n$)
$\subset$ so($2n$) $\subset$ so($2n+1$). The two metric components
can be specified as follows. Consider first the (``standard'')
Cartan-Killing metric $K$ on the entire Lie algebra $ {\cal G}=$
so($2n+1$). We choose the basis of generators $T_i$ such that
$K(T_i,T_j) \propto \delta_{ij}$  (the structure constants with
indices lowered by this metric are then totally antisymmetric). By
restriction this is an $H$-invariant bilinear form on  the
subspace ${\cal G}/{\cal H}$, on which it is block-diagonal on the
two irreducible representation spaces of $H$. The scales of the
metric on the two blocks represent the two parameters of the
metric, say $\eta_1\geq 0$ and $\eta_2\geq 0$, and we can write
explicitly \beq \eta_{I,J} = \eta_1 \delta_{I, I_1} \delta_{J,
J_1} K(T_{I_1}, T_{J_1}) + \eta_2 \delta_{I, I_2} \delta_{J, J_2}
K(T_{I_2}, T_{J_2}). \label{metrictwoparameters} \eeq Note that
one may relate the structure constants $f_{IJK}$ of Eq.\
(\ref{Riemann}), with indices lowered with the metric $\eta_{IJ}$,
to those with indices lowered with the Killing metric $K(T_I,
T_J)$, which are totally antisymmetric.

The computation of the Ricci (and Riemann) tensor of the
homogeneous space so($2n+1$)/u($n$) is tedious but
straightforward, using (\ref{Riemann}), (\ref{Ricci}),
(\ref{metrictwoparameters}). In terms of the following
parametrization of the metric,
\begin{equation}
\eta_1 = {1\over g}, \qquad
 \eta_2 = {1\over 2x g},
\label{parametrizationmetric}
\end{equation}
one obtains from (\ref{betacoset}) the one-loop beta functions:
\begin{eqnarray}
{d g \over d l} &=& 2g^2   [ x^2 + (n-1)]    + O(g^3),
\label{betafunctionN}\\
 { d  x \over dl} &=& 2 (n-1) g x (x-1) [  1-
{n\over n-1} x]   + O(g^2). \label{betafunctionx}
\end{eqnarray}
These equations are valid in the limit $g\to 0$ with $x$ fixed.

The parameter $x\geq 0$ measures the relative strength of the two
metric components. There are two special cases,  $x=0$ and $x=1$,
which we now discuss in turn. Consider the chain of vector spaces
(Lie algebras) u($n$) $\subset$ so($2n$) $\subset$ so($2n+1$). As
$x\to 0$, one sees from Eq.\ (\ref{parametrizationmetric}) that
the stiffness of the fluctuations of the sigma model
(\ref{sigmaaction}) associated with the metric component $\eta_2$,
that is of those in the space so($2n+1$)/so($2n$), becomes
infinite. At $x=0$ these fluctuations in the gradients (with
respect to $r$) of the sigma model field are forbidden, and the
only remaining fluctuations are those associated with the metric
component $\eta_1$, that is of those in so($2n$)/u($n$), together
with a degree of freedom on SO($2n+1$)/SO($2n$) (a sphere,
$S^{2n}$) which is independent of $r$ and is therefore global.
This is related to the structure of SO($2n+1$)/U($n$), which
[because of the chain of subgroups U($n$) $\subset$ SO($2n$)
$\subset$ SO($2n+1$)] can be viewed as a fiber bundle with base
space SO($2n+1$)/SO($2n$)$\cong S^{2n}$, and fiber
SO($2n$)/U($n$). Thus for each point on the sphere $S^{2n}$, there
is a copy of the space SO($2n$)/U($n$) in which the field can
fluctuate locally. Because of the global degree of freedom on
$S^{2n}$, there is still a global SO($2n+1$) symmetry. In simple
terms, the symmetry is spontaneously broken to SO($2n$); this does
not violate the Hohenberg-Mermin-Wagner theorem, which applies for
integer $n>1$, because the coupling $1/\eta_2=0$. Neglecting the
global degree of freedom, the line $x=0$ now corresponds to the
SO($2n$)/U($n$) sigma model. (These remarks explain why only
non-negative powers of $x$ appear in the perturbative beta
functions.) This line is an invariant of the RG flow, and the beta
function (\ref{betafunctionN}) reduces to that of the
SO($2n$)/U($n$) sigma model, which to three-loop order
\cite{hikami} is:
\begin{eqnarray} {dg \over dl}&=&2(n-1) g^2 + 2(n^2 -3 n +4) g^3 \nonumber\\
&&{}+ (3n^3 - 14 n^2 + 35 n-28)g^4 + {\cal O}(g^5).
\label{hikbeta}\end{eqnarray} For $n>1$, $g$ flows to large
values.

At $x=1$, on the other hand, one can check by direct calculation
that the metric in (\ref{metrictwoparameters}) reduces to that of
the symmetric space SO($2n+2$)/U($n+1$) of higher symmetry.
Therefore, the line $x=1$ must also be an invariant of the RG
flow. On this line, the one-loop beta function in
(\ref{betafunctionN}) reduces to that of the symmetric space
SO($2n+2$)/U($n+1$), Eq.\ (\ref{hikbeta}) with $n\mapsto n+1$, as
expected. One sees from (\ref{betafunctionx}) that for $n>1$ both
lines $x=0$ and $x=1$ are attractive at weak coupling. The line
$x=(n-1)/n$ is the separatrix between these two regimes.

There is also a limit, $\eta_1=0$, in which fluctuations in
SO($2n$)/U($n$) are ``soft'', and can be gauged away; compare the
discussion in Ref.\ \onlinecite{subo(4)}. In this case the model
reduces to the nonlinear sigma model with target space
SO($2n+1$)/SO($2n$) $\cong S^{2n}$, mentioned earlier. However,
this strong-coupling limit cannot be accessed perturbatively in
$1/\eta_1$, $1/\eta_2$.

In the replica limit $n \to 0$, of interest in this paper, we
obtain Eqs.\ (\ref{betafunctionNzero}). Note that now, in contrast
to the case $n>1$, $x=0$ is repulsive, and that on the line $x=0$,
$g$ flows towards weak coupling. Near $x=0$, the one-loop flow
lines are hyperbolas, $xg=$ constant. In the vicinity of the line
$x=1$ the one-loop flow lines are exponentials: $g/g^\ast = \exp
\{ - 2(x-1) \}$. The one-loop flows are qualitatively different
depending on the sign of $(x-1)$. When $x<1$ the coupling constant
$g$ decreases upon RG flow until it approaches some asymptote
$g=g^\ast$ at $x=1$, while for $x>1$, on the other hand, $g$
increases towards $g^\ast$ and $x$ decreases towards $1$, as $l\to
\infty$.

To two-loop order, as discussed in Sec. \ref{nlsm}, the region
around $x=1$ flows towards strong coupling. We now consider the
behavior of the flows, in particular those which start with bare
values near $x=0$. Use of the one-loop equations
(\ref{betafunctionNzero}) near $x=0$, with bare values $x_0$ and
$g_0$, with $x_0$ and $g_0x_0$ assumed small, shows that a value
of $x$ of order 1 is reached when $l-l_0$ ($l_0$ is the logarithm
of the short distance cutoff, the scale at which $x_0$, $g_0$ are
defined) is $l-l_0\simeq 1/(2g_0x_0)$, at which $g$ is of order
$g_0x_0$. Then we use the two-loop flows at $x=1$, which should be
sufficient accuracy, starting from these values. Integrating Eq.\
(\ref{hikbetanzero}), we find finally for the crossover that
passes close to the two fixed points at $x=0$ and $x=1$, that $g$
becomes of order one when the length scale $e^l$ is
\beq\exp[1/(2g_0x_0)+1/(8g_0^2x_0^2)],\eeq in units of the short
distance cutoff, $e^{l_0}$ (the numerical factors in the exponent
should not be taken too literally). The two-loop corrections near
$x=0$ will generate only a factor of a power of $g$ in this length
scale. Flows that start at $x>1$ and $g$ small give a similar
scale, $\sim \exp[1/(8g_0^2)]$.


\end{document}